\documentclass[oldversion]{aa}
\usepackage{graphicx}
\usepackage{txfonts}
\usepackage{natbib}
\usepackage{subfigure}
\usepackage{color}
\begin{document}


   \title{Simulations of galactic disks including a dark baryonic component}

   \author{Y. Revaz\inst{1,2} 
          \and 
	  D. Pfenniger\inst{3}
          \and 	
	  F. Combes\inst{2}
	  \and
	  F. Bournaud\inst{4}  
}

   \offprints{Y. Revaz}

   \institute{Laboratoire d'Astrophysique, \'Ecole Polytechnique F\'ed\'erale de Lausanne (EPFL), 1290 Sauverny, Switzerland\\
              \and
              LERMA, Observatoire de Paris, 61 av. de l'Observatoire, 75014 Paris, France\\
              \and
              Geneva Observatory, University of Geneva, CH-1290 Sauverny, Switzerland\\
              \and
              Laboratoire AIM, CEA-Saclay DSM/DAPNIA/SAp-CNRS-Universit\'e Paris Diderot, 91191 Gif-Sur-Yvette\\
}

   \date{Received -- -- 20--/ Accepted -- -- 20--}

 
  \abstract
{
The near proportionality between HI and dark matter in outer galactic
disks prompted us to run N-body simulations of galactic disks in which
the observed gas content is supplemented by a dark gas component
representing between zero and five times the visible gas content.
While adding baryons in the disk of galaxies may solve some issues, it
poses the problem of disk stability.  
We show that the global stability is ensured if the ISM is multiphased, composed of two
partially coupled phases, a visible warm gas phase and a weakly
collisionless cold dark phase corresponding to a fraction of the
unseen baryons.  The phases are subject to stellar and UV background
heating and gas cooling, and their transformation into each other is
studied as a function of the coupling strength.  
This new model, which still possesses a dark matter halo, fits
the rotation curves as well as the classical CDM halos, 
but is the only one to explain the existence of an open and contrasting spiral structure, 
as observed in the outer HI disks
}


   \maketitle

%

\section{Introduction}

The $\Lambda$CDM scenario encounters much success in reproducing
the large scale structures of the Universe traced by the
Lyman-$\alpha$ forest and gravitational lensing \citep{springel06}.
However, at galactic scale, where the baryonic physics plays a major
role, the $\Lambda$CDM scenario has several well known
problems.  The dark matter cusp in galaxies 
\citep{blaisouellette01,deblok02,swaters03,gentile04,gentile05,spekkens05,deblock05,deblok08,spano08}, and in
particular in dwarf irregulars, the high number of small systems
orbiting halos \citep{moore99,klypin99,stringari07}, the low angular
momentum problem at the origin of too small disks
\citep{navarro91,navarro97,kaufmann07}, as well as the difficulty of
forming bulgeless disks \citep{mayer08} suggest that some physics is
missing.

A strong prediction of the $\Lambda$CDM scenario is that the so called
``missing baryons'' reside in a warm-hot gas phase in the over-dense
cosmic filaments \citep{cen99,cen06}.  However, there are now several
theoretical and observational arguments that support the fact that
galactic disks may be more massive than usually thought, containing a
substantial fraction of these ``missing baryons".

It has been pointed out by \citet{bosma78,bosma81a} that in
samples of galaxies, the ratio between the dark matter and HI surface
density is roughly constant well after the optical disk \citep[see
  also][]{carignan90,broeils92, hoekstra01}.  This correlation may be
a direct consequence of the conservation of the specific angular
momentum of the gas during the galaxy formation process
\citep{seiden84}.  However, as shown by \citet{van_den_bosch01} the
angular momentum conservation leads to the formation of too
concentrated stellar disks, a real problem for the low surface
brightness galaxies.  On the other hand, the HI-dark matter
correlation may suggest that a large fraction of the dark matter lies
in the disk of galaxies, following the HI distribution.  This physical
link between HI and dark matter has been confirmed by the baryonic
Tully-Fischer relation of spiral galaxies \citep{pfenniger05} and more
recently for extremely low mass galaxies \citep{begum08}.

Dark matter in the disk of galaxies is now also suggested by dynamical
arguments based on the asymmetries of galaxies, either in the plane of
the disks or transverse to it.  For example, the large spiral
structure present in the very extended HI disk of NGC\,2915 is
supported by a quasi self-gravitating disk \citep{bureau99,masset03}.
\citet{revaz04} have shown that heavy disks are subject to vertical
instabilities (also called bending instabilities) and may generate all
types of observed warps: the common S-shaped but also the U-shaped and
asymmetric ones \citep{reshetnikov99,sanchez-saavedra03}. Other
arguments for the origin of warps have been proposed
\citep{jiang99,shen06,weinberg06}, however none of them are able to
explain the three types of warps.

Several works have tried to study the flattening of the Milky Way halo
using potential tracers like dwarf galaxies \citep{johnston05}. From
theses studies, the halo potential appears to be nearly spherical,
with an ellipticity of $0.9$. However, dwarf galaxies trace the halo
at distances larger than $20\,\rm{kpc}$ where it is difficult to
distinguish ``classical'' CDM disks with a slightly flattened halo
from a heavy disk model, containing half of the halo mass in an
extended flat disk \citep{revaz07}.

A more accurate method is to trace the potential near the plane of the
disk, using for example the vertical gas distribution.
\citet{kalberla03,kalberla04,kalberla07} have modeled the HI
distribution out to a galactic radius of $40\,\rm{kpc}$.  Their
self-consistent model is compatible with a self-gravitating dark
matter disk having a mass of $2-3\,\times 10^{11}\,\rm{M_\odot}$.

The presence of dark baryons in the disk of galaxies is reinforced by
the numerous signs of recent star formation in the far outer disk of
galaxies, correlated with HI gas in NGC\,6946 \citep{ferguson98},
NGC\,628 \citep{lelievre00}, M\,101 \citet{smith00}, M\,31
\citet{cuillandre01}, NGC\,6822 \citep{deblok03}, M\,83
\citep{thilker05}.  Correlations between young stars and HI far from
the center reveal that molecular gas, at the origin of a weak but
existing star formation rate, must be present in abundance there,
despite the lack of CO detection.

Other arguments supporting the idea that dark matter could reside in
the galactic disk in the form of cold molecular hydrogen ($\rm{H}_2$) have
been widely discussed by \citet{pfenniger94a}.

Unfortunately, a direct detection of cold $\rm{H}_2$ in the outer disk of
galaxies appears to be a very hard task \citep{combes97}.  As the
$\rm{H}_2$ molecule is symmetrical, any electric dipole moment is
canceled.  The molecule may be detected in emission only by its
quadrupole radiation, the lowest corresponding to a temperature of
$512\,\rm{K}$ above the fundamental state, much above the
$5-7\,\rm{K}$ expected from \citet{pfenniger94b}.  A weak radiation of
the $\rm{H}_2$ ultra-fine lines is expected at kilometer wavelengths.
However, since it is $10$ orders of magnitude smaller than the HI line
for the same density, its detection will not be possible in the
near future.  The absorption lines may be the best way to detect
$\rm{H}_2$. But this method requires large statistics, as the filling
factor of the gas is expected to be very low ($<1\%$).

The best tracers of cold $\rm{H}_2$ may be the pure
rotational lines of $\rm{H}_2$ (at 28, 17, 12 and 9 microns), which
could be emitted by a few percent of the molecular gas, excited by
intermittent turbulence \citep[see for example][]{boulanger08}.

Indirect detection by tracers may be prone to error.  For example, CO
traces the $\rm{H}_2$ but only for enriched gas and fails at large
distances from galaxy centers.  Moreover, it is impossible to detect
CO emission from a cloud at a temperature close to the background
temperature.  The cold dust component detected by COBE/IRAS
\citep{reach95} is known to trace the cold $\rm{H}_2$.  But as for the
CO, it is limited to small galactic radii where the cold gas is still
mixed with some dust.  Spitzer mid-infrared observations have recently
revealed that large quantities of molecular hydrogen are not
associated with star formation \citep{appleton06,ogle07}. An
unexpectedly large amount of $10^{10}\,\rm{M_\odot}$ of H$_2$ is revealed
only by its strong H$_2$ emission lines in the galaxy Zwicky 3146
\citep{egami06}.  \citet{grenier05} showed that much gas in the solar
neighborhood is revealed only by $\gamma$-rays.  Other indirect
detections could be possible using micro-lensing events
\citep{draine98,fux05}. However, this method has not been exploited up
to now.

As direct and indirect observational detection of cold gas is
difficult, it is necessary to test the effect a cold gas component
would have on the global evolution of galactic disks.  In this paper,
we present new N-body simulations of galactic disks, where the
observed gas content has been multiplied by a factor between 3 and
5. In addition to this extra dark baryonic component, a non-baryonic,
spheroidal pressure-supported dark halo containing most of the
large-scale dark mass is conserved.

An important issue that our model aims to answer is the stability
question of heavy disks \citep{elmegreen97}.  Our model assumes that
the additional baryons lie in a very cold and clumpy phase
\citep{pfenniger94b}, partially dynamically decoupled from the visible
dissipative phase.  We show that this phase can thus be
less dissipative than the visible ISM and has larger velocity
dispersions, so that the global disk stability is preserved.  A new
numerical implementation of the cycling acting between these two
phases is proposed.


The secular evolution shows that the models with additional baryons
are globally stable and share on average the same observational
properties as the ``classical'' CDM disks. However, they give a
natural explanation for the presence of contrasting spiral structures in
the outskirts of HI disks which is difficult to explain when
taking into account the self-gravity of the HI alone.  In a forthcoming
paper, we will show that this model also reproduces the puzzling dark
matter content present in debris from galaxies \citep{bournaud07}.

Our model is different from previous multiphase models
\citep{semelin02,harfst06,merlin07} in the sense that it does not
compute a cycling between a cold-warm dissipative and hot medium, but
between a very cold weakly collisional phase and the visible
dissipative phase.

The paper is organized as follows. Details of the multiphase model is
given in Section~\ref{multiphase_model}.  In Section~\ref{parameters}
we briefly discuss the parameters used, and Section~\ref{galaxy_model}
is devoted to the galaxy model
description. Section~\ref{model_evolution} compares the evolution of
galaxy models with and without additional baryons and a short
discussion. A summary is given in Section~\ref{conclusions}.


\section{The multiphase model}\label{multiphase_model}


\subsection{The straightforward approach}

The straightforward approach when modelling the galactic ISM is to
assume that gas behaves like an ideal, inviscid gas.  The evolution of
the specific energy $u$ of the gas may be obtained by inserting the
continuity equation into the first law of thermodynamics:
	\begin{equation}
	\frac{du}{dt}=  \frac{P}{\rho^2}\frac{d\rho}{dt} + \frac{\Gamma(\rho) - \Lambda(\rho,T)}{\rho}. 
	\label{dudt}
	\end{equation}
The first right hand side term corresponds to the adiabatic behavior
of the gas, while the second is responsible for the entropy
variation. This latter reflects the non-adiabatic processes included
through the heating and cooling function $\Gamma(\rho)$ and
$\Lambda(\rho,T)$.  At low temperature ISM ($T<10^4\,\rm{K}$), the
heating is dominated by the cosmic ray heating \citep{goldsmith69} and
by the photo-electric UV heating on grains \citep{watson72,draine78}.
As soon as the gas is enriched with metals, the cooling function is
dominated by the CII, SiII, OI, FeII line emissions \citep{maio07}.  Since
the cooling function depends on the square of the gas density, the gas
temperature is sensitive to it.

However, the ISM is known to be strongly non homogeneous down to very
small scales, reaching densities higher than $10^6\,\rm{atoms\times
  cm^{-3}}$ where the cooling time is very short, leading to
equilibrium temperatures below $10\,\rm{K}$.  Such over-densities are
unfortunately far from being resolved by galactic scale simulations
and numerical simulations miss the associated low temperatures.
While being physically correct, Eq.~\ref{dudt} will then strongly bias
the equilibrium temperature of the gas, because the estimated average
density $\rho$ poorly reflects the actual physics.

\subsection{Statistical approach}

\begin{figure}
  \input{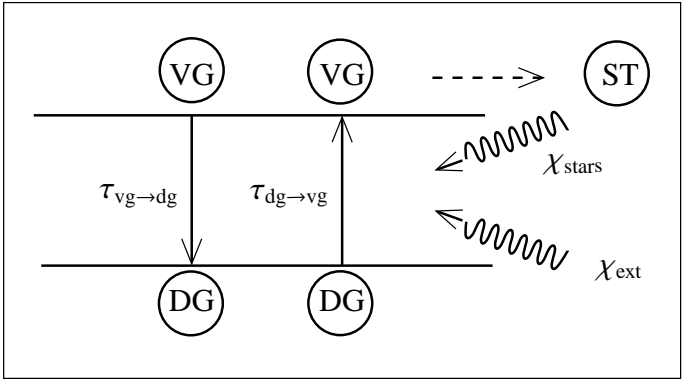}
  \caption{Schematic representation of the ISM two levels system. The
    visible gas is labeled VG, the dark gas DG and the stars ST. The
    probability of transition between the visible and the dark gas
    depends on the UV flux generated by young stars ($\chi_\star$) and
    from extragalactic sources ($\chi_{\rm{ext}}$) setting the
    characteristic transition times $\tau_{\rm{vg} \to \rm{dg}}$ and
    $\tau_{\rm{dg} \to \rm{vg}}$.  Stars are formed out of the visible
    gas only.  }
  \label{two_levels}

  \end{figure}

Instead of following the thermal specific energy of particles using
the biased Eq.~\ref{dudt}, we propose a new statistical approach
avoiding the problem of the density and temperature evaluation.  The
multiphase ISM is assumed to be a two level system (see
Fig.~\ref{two_levels}) with probabilities of transition depending only
on the local excitation energy flux, that we call for short ``UV
flux'', which is assumed to be the dominant heating process.

The top level is populated by the well known observed dissipative gas
detected by its CO, H$_2$ or HI emission.  This phase will be called
the visible gas.  The bottom level is populated by undetected very
clumpy and cold gas as proposed by \citet{pfenniger94b}, having
temperature below the CO detection limit, at temperature equilibrium
with the cosmic background radiation.  This gas results from the
strong cooling that occurs in overdense regions.  As this gas is
missed by all tracers, we will call it the dark gas.

According to the astrophysical literature, we have used here the word
``gas''.  However, it is well known that the cold interstellar medium
shows fractal properties which have been observed up to the instrumental
capabilities, down to a few hundred AU \citep{heithausen04}.  Such
heterogeneous fluids clearly do not have the viscosity or
other mean properties of smooth flows.  In other fields, it would be
called granular flows, for example.
We are aware of the degree of simplification of our model compared to the
complexity of the ISM.  A more complex model should include
the dark component, the CO-undetected metal-poor warmer H$_2$ gas that
may exist in the outskirts of galactic disks \citep{papadopoulos02},
and possible effects related to phase transition and separation in
the He-H$_2$ mixture at very cold temperature \citep{safa08}.
%

The transition times between visible and dark gas is only dependent on
the local UV flux $\chi$, which is the main heating process in the
ISM. For the transition times, we have chosen the following simple
relations:
	\begin{equation}
	\left\{ 
	\begin{array}{lll}
	\tau_{\rm{vg} \to \rm{dg}} &=& \chi^{+\beta}\tau, \\
	\tau_{\rm{dg} \to \rm{vg}} &=& \chi^{-\beta}\tau, \\
	\end{array}
	\right.
	\end{equation}
where $\beta$ and $\tau$ are two free parameters. $\tau$ gives the
time scale of the transition, independently of the outer flux. $\beta$
sets the flux dependency. For $\beta=0$, the transition does not depend
on the outer flux.  At equilibrium, these equations provide
simple relations for the ratio between the visible gas density
$n_{\rm{vg}}$, the dark gas density $n_{\rm{dg}}$ and the total gas
density $n_{\rm{tg}}$, as a function of the normalized local UV flux
$\chi$ only
\footnote{The solution is the one of a two level transition system with
probabilities $p_{\rm{wc}}\sim 1/\tau_{wc}$ and $p_{\rm{cw}}\sim 1/\tau_{cw}$.}:
	\begin{eqnarray}
	\label{vgtg1}
	\frac{n_{\rm{vg}}}{n_{\rm{tg}}}	&=&  \frac{\chi^{2\beta}}{1+\chi^{2\beta}} = \gamma,\\
	\label{vgtg2}
        \frac{n_{\rm{dg}}}{n_{\rm{tg}}}	&=&  \frac{1}{1+\chi^{2\beta}} = 1-\gamma,\\
	\label{vgtg3}
	\frac{n_{\rm{dg}}}{n_{\rm{vg}}} &=&  \chi^{-2\beta},
	\end{eqnarray}
where in these latter equations, we have defined $\gamma$ as
$n_{\rm{vg}}/n_{\rm{tg}}$.  The interpretation of these equations is
the following.  When the gas is subject to a strong UV field, the
heating dominates and the probability of transition between the dark
and the visible gas is high. The gas is then dominated by the visible
phase.  On the contrary, when the UV field is very low (as the
case in the outer regions of the galactic disk), the cooling dominates
and the probability of a transition between the dark and the visible gas
is low. The gas is then dominated by the dark phase.

\subsection{Visible and dark gas dynamics}\label{visible_and_dark}

The visible phase is assumed to be dissipative, because it is more diffuse
and collisional than the cold phase.  Instead of using the classical
SPH approach, which makes the gas strongly collisional, we have
preferred to use the sticky particle scheme
\citep{brahic77,schwarz81,combes84} which better simulates the clumpy
ISM.  In the following simulations, we have used $\beta _{\rm r}=0.8$
and $\beta _{\rm t}=1$ in order to strictly conserve angular
momentum\footnote{see \citet{bournaud02} for details on the meaning of
  $\beta _{\rm r}$ and $\beta _{\rm t}$.}.  The frequency of
collisions between particles is set to be proportional to the local
visible gas density.

Also, as the very cold and clumpy dark gas does not radiate, it is
expected to be weakly collisional.  Its relaxation time being much
longer than the dynamical time, we neglect here the effect of
collisions.


\subsection{Star formation}

Stars that are assumed to be the main source of UV flux in the inner
part of the galaxy may be formed out of the visible gas only. We have
used a classical star formation recipe \citep{katz96} that reproduces
well the Schmidt law.

\subsection{UV flux}

The normalized UV energy flux $\chi$ is decomposed into two parts. A
stellar radiation flux $\chi_\star$ and an extragalactic background
energy flux $\chi_{\rm{ext}}$.

\subsubsection{Stellar UV flux}

The stellar normalized radiation flux is computed by summing the
contribution of each star, assuming an appropriate $(L/M)$ ratio.
	\begin{equation}
	\chi_\star =  \frac{1}{4\pi c U_{\odot}} \sum \frac{m_i (L/M)_i}{r_i^2 + \epsilon^2},   
	\label{chi_star}
	\end{equation}
where $m_i$ is the particle of mass i and $r_i$ its distance to the
point where the flux is estimated.  $\epsilon$ is the gravitational
softening, $c$ the speed of light and $U_{\odot}$ is the Habing UV
energy density equal to $5.4\times 10^{-14}\,\rm{erg/cm^3}$
\citep{habing68}.  The ratio $(L/M)$ is set to $1.58\times
10^{31}\,\rm{erg/s/M_\odot}$. This value is constant for all particles
(independently of the mass and type of star particles).
The summation in Eq.~\ref{chi_star} is performed over all star
particles and may be very time consuming in a large N-body system.
We have computed it by taking advantage of its similarity to the
summation of gravitational forces which is computed using the
\emph{treecode} method. The time overhead for the computation of
$\chi_\star$ is then negligible.

\subsubsection{Extragalactic UV flux}

The extragalactic UV background normalized radiation flux is assumed
to be constant.
	\begin{equation}
	\chi_{\rm{ext}} =  \frac{U_{\rm{ext}}}{U_{\odot}}
	\end{equation}
where the extragalactic UV background density is set to $2\times
10^{-15}\,\rm{erg/cm^3}$.

%
%
%
%
%


\section{Parameters}\label{parameters}

\begin{figure}
  \resizebox{\hsize}{!}{\includegraphics[angle=0]{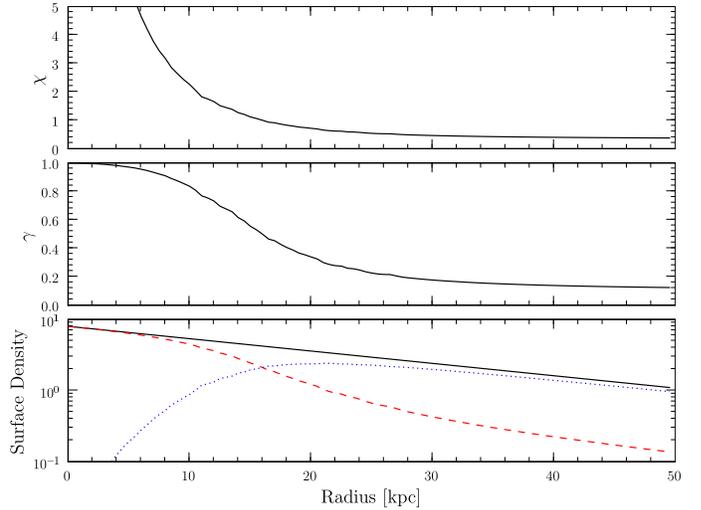}}
  \caption{Normalized UV flux $\chi$, visible to total gas density
    ratio $\gamma$ and surface density of the gas, as a function of
    the galactic radius $R$. In the bottom graph, the red dashed line
    corresponds to the visible gas, the blue dotted line to the dark
    gas and the dark plain line to the total gas.}
  \label{X-gamma-n-nw-nc}
  \end{figure}

Our multiphase model is based on 4 parameters.  The external energy
density flux $U_{\rm{ext}}$, the UV light to mass ratio $(L/M)$, the
coefficient parameter $\beta$ and the time scale of transition $\tau$.
In this paper, we have fixed the 4 parameters to constant values given
in Table~\ref{multiphase_parameters}.
%
\begin{table}
    \begin{tabular}{c c c c}
    \hline
    \hline
    $\beta$   & $(L/M)$                            & $U_{\rm{ext}}$             & $\tau$  \\
              & $[\rm{erg}/\rm{s}/\rm{M_\odot}]$   & $[10^{11}\,\rm{M_\odot}]$  & $[\rm{Gyr}]$ \\
    \hline
    \hline
     $1$        & $1.58\times 10^{32}$              & $1.8\times 10^{14}$         & 10\\
    \hline
    \end{tabular}
    \caption[]{Multiphase model parameters.}
    \label{multiphase_parameters}
\end{table}
The parameter $\beta$ is set to 1 for simplicity. The
parameter $(L/M)$ is chosen such that the normalized stellar UV flux
$\chi_\star$ is unity near $15\,\rm{kpc}$ when the stellar
distribution at the origin of the UV flux corresponds to a realistic
exponential disk, as will be presented in
Section~\ref{galaxy_model}.  According to Eq.~\ref{vgtg3}, these
values correspond to the radius where the dark component surface
density is equivalent to the visible surface density.  $U_{\rm{ext}}$
is chosen such that, in the absence of a stellar UV field (at a distance
$R=\infty$ from the galaxy center), the ratio
$n_{\rm{vg}}/n_{\rm{vg}}=\gamma_\infty$ where $\gamma_\infty$ has been
set to $1/10$.  From Eq.~\ref{vgtg1} and setting $\chi_\star=0$ we
get:
	\begin{equation}
	U_{\rm{ext}} = U_{\odot}\left( \frac{\gamma_\infty}{1-\gamma_\infty}  \right)^{\frac{1}{2\beta}}.
	\end{equation}
The time scale of the transition $\tau$ is set to about $10$ dynamical
times so that the cycling between visible and dark gas is slow.
Fig.~\ref{X-gamma-n-nw-nc} shows the total normalized flux $\chi$ as
well as the visible over total gas mass ratio $\gamma$ obtained with
the parameters of Table~\ref{multiphase_parameters} and the stellar
distribution of Section~\ref{galaxy_model}.  As an illustration, the
flux is applied on a purely exponential baryonic gaseous disk.  The
resulting surface densities for the visible and dark gaseous disk are
displayed in the bottom of the figure.  The exponential disk generates
a strong UV flux at the center. Consequently, the visible gas
dominates there. On the contrary, in the outer part, the UV flux is
dominated by the constant weak extragalactic flux, making the dark gas
dominant.
%


\section{Galaxy models}\label{galaxy_model}


We have used our multiphase model to simulate the evolution of Milky
Way-like spiral galaxies.  In order to understand the effect of the
multiphase model, we have compared two models, including additional
dark baryons, with a reference ``classical'' model. In addition, we
have build a fourth simulation including the perturbation of
$\Lambda$CDM substructures.  In the following sections, we first
present our reference model (also called $N=1$ model), where no
additional dark baryons have been added.  In the next section (Section
\ref{dark_baryons_model}) we present how it is possible to add
dark baryons in the reference model, in a way consistent with the
rules proposed in Chapter~\ref{multiphase_model} and by conserving the
same observational properties of the reference model.  The model
including $\Lambda$CDM substructures will be presented in
Section~\ref{LCDM substructures}.  Section \ref{velocity_distribution}
will discuss how initial velocities have been set in order to make the
disk stable.


\subsection{Mass distribution}

\subsubsection{The reference model}\label{reference_model}

Our reference model is designed to fit typical properties of giant Sbc
galaxies with a flat extended rotation curve.  It is initially composed
of a bulge, an exponential stellar disk, a gas disk and a dark matter
halo.

\begin{figure}
  \resizebox{\hsize}{!}{\includegraphics[angle=0]{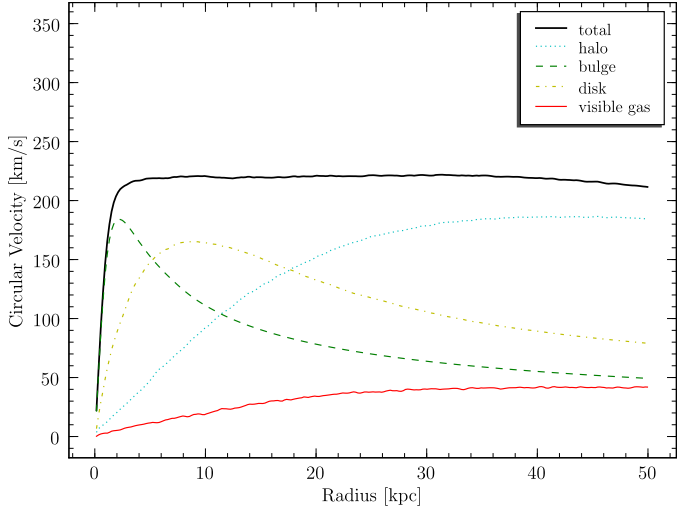}}
  \caption{Rotation curve of the reference model with the contribution
    of each component, the bulge, the exponential disk, the visible
    gas and the dark matter halo.}
  \label{crv_ref}
  \end{figure}

\begin{enumerate}
\item  
The bulge density profile is a simple Plummer model:
\begin{equation}
  \rho^{\rm{b}}(R,z) \propto {\left( 1+ \frac{r^2}{r_{\rm{b}}^2} \right)^{-5/2}},
   \label{bulge_density}
\end{equation}
with a characteristic radius $r_{\rm{b}}=1.4\,\rm{kpc}$.

\item
The exponential stellar disk takes the usual form:
\begin{equation}
  \rho^{\rm{d}}(R,z) \propto e^{-R/H_R}\,e^{-|z|/H_z} ,
   \label{disk_density}
\end{equation}
where the radial and vertical scale length are
respectively $H_R=4$ and $H_z=0.3\,\rm{kpc}$.

\item
The dark matter is distributed in a Plummer model:
\begin{equation}
  \rho^{\rm{h}}(R,z) \propto  \rho^{\rm{MN}}_{(0,r_{\rm{dm}})},
  \label{halo_density}
\end{equation}
with a characteristic radius $r_{\rm{dm}}=30\,\rm{kpc}$, ensuring a
flat rotation curve up to $40\,\rm{kpc}$.  The model is truncated at
$3\,r_{\rm{dm}}$ ($90\,\rm{kpc}$).  In order to avoid any perturbation
of the halo on the disk due to imperfect equilibrium, we have set it
as a rigid potential. We have checked that a live halo will not
influence our results.

\item
The choice of the visible gas distribution follows the observations of
\citet{hoekstra01}, where the dark matter contribution to the rotation
curve is a multiple of the contribution of the gas.  This is achieved
by distributing the gas in a Miyamoto-Nagai model \citep{miyamoto75}:
\begin{equation}
  \rho^{\rm{vg}}(R,z) \propto \rho^{\rm{MN}}_{(h_z,r_{\rm{dm}})}.
  \label{gas_density}
\end{equation}
The proportionality between the Plummer halo (being in fact a subclass
of a Miyamoto-Nagai model) and the disk is ensured if the disk scale
length is similar to the one of the halo : $h_R=r_{\rm{dm}}$ (see
Appendix~\ref{appendix1}).  The vertical scale height $h_z$ is fixed
to $0.3\,\rm{kpc}$.  We also have included a flaring of the visible
gas disk by multiplying the $z$ coordinates of the particles by:
\begin{equation}
  \exp{(\frac{R}{R_{\rm{f}}})},
  \label{flaring}
\end{equation}
where the transition radius $R_{\rm{f}}$ is set to $40\,\rm{kpc}$.

\end{enumerate}
Mass and parameters of the model are summarised in
Table~\ref{mass_model} and the corresponding rotation curve is
displayed in Fig.~\ref{crv_ref}.

\begin{table}
    \begin{tabular}{l l c c}
    \hline
    \hline
    component  &  model & parameters  & mass  \\
               &        &  $[\rm{kpc}]$ &  $[10^{11}\,\rm{M_\odot}]$ \\
    \hline
    \hline
    bulge	& Plummer	  & $r_{\rm{b}}=1.4$	& 0.28\\
    disk	& exponential	  & $H_R=4$,$H_z=0.3$   & 0.69\\
    halo	& Plummer	  & $r_{\rm{dm}}=30$    & 5.17\\
    gas	        & Miyamoto-Nagai  & $a=30$,$b=0.3$      & 0.21\\
    \hline
    \end{tabular}
    \caption[]{Parameters for the reference giant Sbc model.}
    \label{mass_model}
\end{table}

\subsubsection{Adding dark baryons in the disk}\label{dark_baryons_model}

Here, we present a method that allows us to build new galaxy models
containing additional dark baryons, but presenting similar
observational properties as the reference model.  The new models are
build by following three rules:
\begin{enumerate}
\item The total rotation curve remains nearly unchanged with respect
  to the reference model.
\item The surface density of the visible gas remains nearly unchanged
  with respect to the reference model.
\item No dynamically significant dark matter is added in the central
  regions.
\end{enumerate}
In addition to these rules, the model, must also respect the ratio
between the visible and additional dark gas that follows from the
multiphase model proposed in Section~\ref{multiphase_model}.

The third rule is derived from the luminosity Milky-Way model
\citep{bissantz02} combined with MACHO micro-lensing observations (see
\citet{gerhard06}, and references therein for more details).  We thus
assume 1) no dark gas exists at the center of galaxies, and 2) the
density of the total gas (tg) is equal to that of the visible gas
(vg):
\begin{equation}
  \rho^{\rm{tg}}_{\rm{in}} = \rho^{\rm{vg}},\, \textrm{in the central regions}.
  \label{central_baryon_density}
\end{equation}
This equation also ensures the second rule at the center.

In the outer part, the total gas mass distribution is constrained by
the first rule.  We can easily transfer mass from the halo to the disk
without changing the rotation curve, using the equivalence in term of
density of Eq.~\ref{mn_pot_sum} (via the Poisson equation).  In
that case, the density of the halo plus the total gas (htg) (including
visible and dark gas) may be written as:
\begin{equation}
  \rho^{\rm{htg}}_{M_{\rm{htg}}} = f \rho^{\rm{MN}}_{(M_{\rm{htg}},h_z,r_{\rm{dm}})} + (1-f) \rho^{\rm{MN}}_{(M_{\rm{htg}},0,r_{\rm{dm}})},
  \label{mn_rho_sum}
\end{equation}
where the first left side term corresponds to the total gas density:
       \begin{equation}
       \rho^{\rm{tg}}_{M_{\rm{tg}},\rm{out}} =      f \rho^{\rm{MN}}_{(M_{\rm{htg}},h_z,r_{\rm{dm}})}, 
       \label{rho_tg}
       \end{equation}
and the second left side term corresponds to the halo density:
       \begin{equation}
       \rho^{\rm{h}}_{M_{\rm{h}}}  =  (1-f) \rho^{\rm{MN}}_{(M_{\rm{htg}},0,r_{\rm{dm}})}.
       \label{rho_h}
       \end{equation}
In this equation, the mass of each component is given as
an index. $M_{\rm{htg}}$ is the sum of the halo and gas mass of the
reference model. $M_{\rm{tg}} = f\,M_{\rm{htg}}$ is the total gas mass
and $M_{\rm{h}} = (1-f)\,M_{\rm{htg}}$ is the halo mass.  It is
convenient to define $f$ as:
\begin{equation}
  f = N\frac{M_{\rm{g}}}{M_{\rm{htg}}},
\end{equation}
where $N$ gives the ratio between the total gas and the gas of the
reference model.  Its value is restricted to the range
$[0,M_{\rm{htg}}/M_{\rm{g}}]$.  When $N=1$, Eq.~\ref{rho_tg} and
\ref{rho_h} give the densities corresponding to the reference model.

We can derive the total gas density all along the disk by combining
Eq.~\ref{central_baryon_density} and \ref{rho_tg}:
\begin{eqnarray}
  \rho^{\rm{tg}}(R) &=& \alpha(R)\,\rho^{\rm{tg}}_{\rm{in}}(R) + \left(1-\alpha(R)\right)\,\rho^{\rm{tg}}_{\rm{out}}(R)   \\
                    &=& \alpha(R)\, \rho^{\rm{vg}}(R) + \left(1-\alpha(R)\right)\,f\,\rho^{\rm{MN}}_{(M_{\rm{htg}},h_z,r_{\rm{dm}})}(R),  
  \label{baryon_density}
\end{eqnarray}
where we have introduced a new function $\alpha(R)$ taking values of
$1$ at the center and $0$ in the outer part of the disk.

We now introduce the multiphase model of
Section~\ref{multiphase_model} that places constraints on the ratio
between visible and dark gas.  Applying Eq.~\ref{vgtg2} to the total
gas density $\rho_{\rm{tg}}$ we can derive the density of the visible
gas:
%
\begin{equation}
  \rho^{\rm{vg}}(R) = \gamma(R)\, \rho^{\rm{tg}}(R)\quad \rm{where} \quad \gamma(R)=\frac{\chi^{2\beta}}{1+\chi^{2\beta}},
  \label{vg_b}
\end{equation}
and where $\gamma(R)$ (see Eq.~\ref{vgtg2}) follows from the stellar
mass distribution, the $(L/M)$ ratio and the UV background.  Combining
Eq.~\ref{baryon_density} and \ref{vg_b} allows us to determine the
value of $\alpha(R)$ which completely defines the total gas
distribution:
\begin{equation}
  \alpha(R) = \frac{1}{\gamma(R)} \left(
  \frac{\gamma(R)\,f\,\rho^{\rm{MN}}_{(M_{\rm{htg}},h_z,r_{\rm{dm}})}(R)-\rho^{\rm{vg}}(R)}{f\,\rho^{\rm{MN}}_{(M_{\rm{htg}},h_z,r_{\rm{dm}})}(R)-\rho^{\rm{vg}}(R)}\right).
  \label{a_R}
\end{equation}
In this equation, we force values of $\alpha$ to stay in the
range $[0,1]$.  As we will see later, this restriction breaks the
second rule in the outer part of the disk.  Finally, the same flaring
as for the visible gas (Eq.~\ref{flaring}) is applied to the total
gas.

Following this scheme, we have constructed two different models,
with respectively $N=3$ and $N=5$.  Applying the multiphase model
with parameters listed in Table~\ref{multiphase_parameters}, we can now
determine the mass properties of the different components (visible
gas, dark gas, total baryons and dark halo) in these models, including
the reference model ($N=1$), given in Table~\ref{N135}.

A larger value of $N$ increases the dark mass and inversely decreases
the dark halo mass (the total mass remaining constant).  The baryon
fraction ($M_{\rm{b}}$/$M_{\rm{tot}}$) is then an increasing function
of $N$ and grows from the universal baryonic fraction up to $30$\% for
model $N=5$. The total visible mass is not strictly identical between
the three models.  Its variation is mainly due to the the outer part
where its surface density decreases with respect to the reference
model. However, the visible gas lying below $20\,\rm{kpc}$ is similar
for the 3 models ($M_{\rm{vg}}=5.4\times\,\rm{10^{9}\rm{M_\odot}}$).

\begin{figure}
  \resizebox{\hsize}{!}{\includegraphics[angle=0]{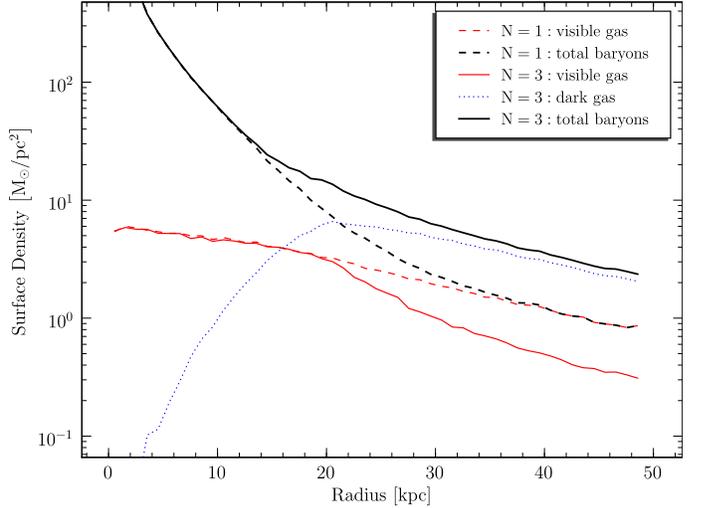}}
  \caption{Comparison of the surface density of baryons between model
    $N=1$ and model $N=3$. The total baryons (including stars) are
    traced in black. The red line represents the visible gas while
    the blue line (only present for the $N=3$ model) falling towards
    the center corresponds to the dark gas.}
  \label{den_003}
  \end{figure}

\begin{figure}
  \resizebox{\hsize}{!}{\includegraphics[angle=0]{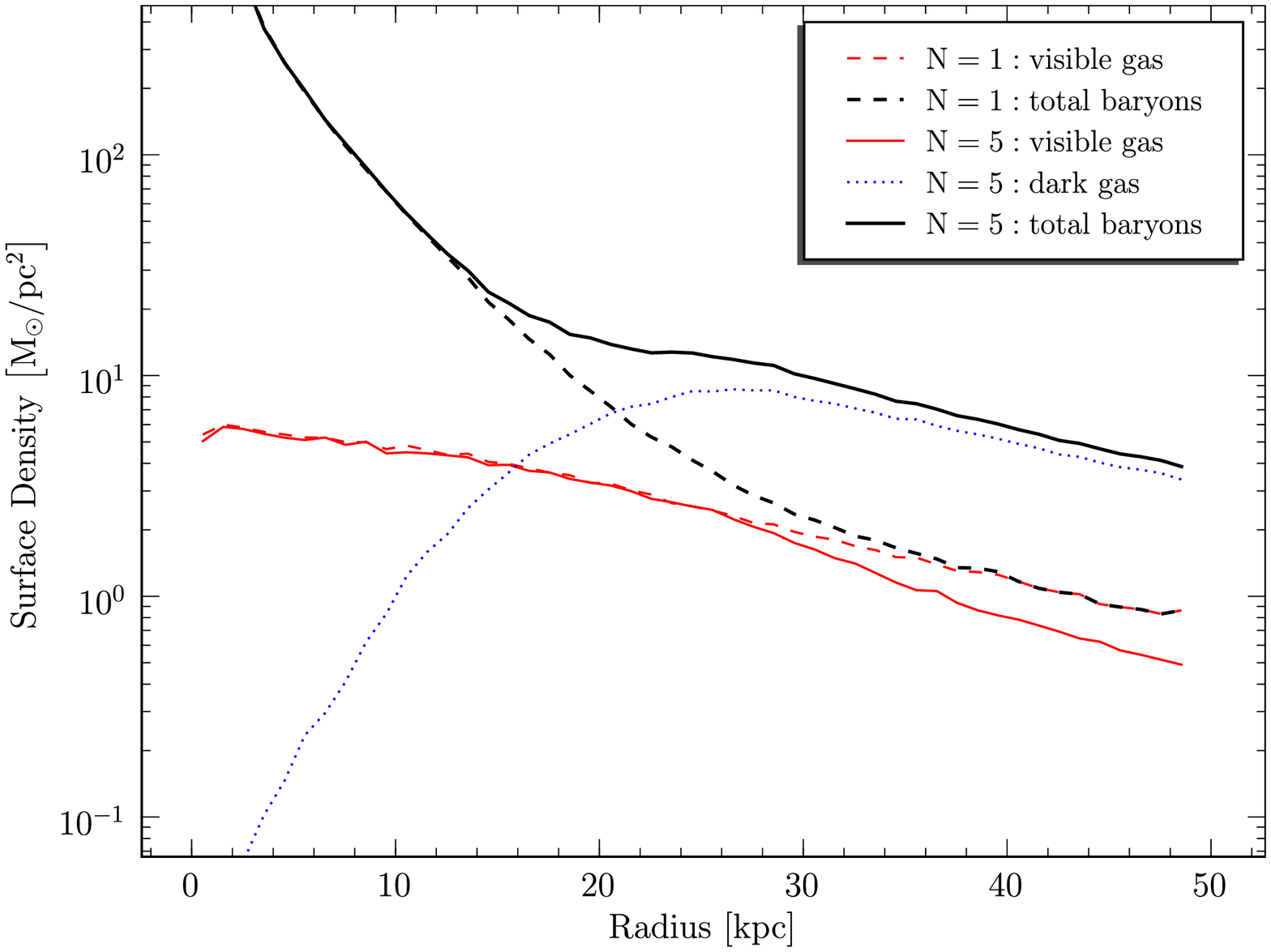}}
  \caption{Same figure as Fig.~\ref{den_003} but for model $N=5$.}
  \label{den_005}
  \end{figure}

The gas and total baryon surface density is displayed in
Fig.~\ref{den_003} and \ref{den_005}.  The outer regions
($R>20\,\rm{kpc}$) of models $N=3$ and $N=5$ are characterised by the
presence of the dark gas which dominates the surface density of the
baryons. In the far outer parts, the baryons are multiplied by $N$,
with respect to the reference model.  On the contrary, due to the drop
in the dark gas surface density, the visible gas surface density of
the inner regions is left unchanged with respect to the reference
model (second rule). The presence of dark gas is only marginal inside
$10\,\rm{kpc}$.

\begin{figure}
  \resizebox{\hsize}{!}{\includegraphics[angle=0]{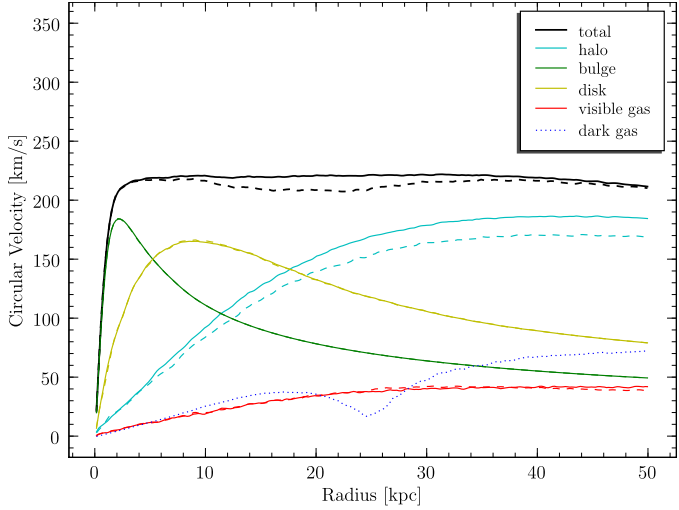}}
  \caption{Comparison of the rotation curves of model $N=1$ and model
    $N=3$ and the contribution of each component.)}
  \label{crv_003}
  \end{figure}

\begin{figure}
  \resizebox{\hsize}{!}{\includegraphics[angle=0]{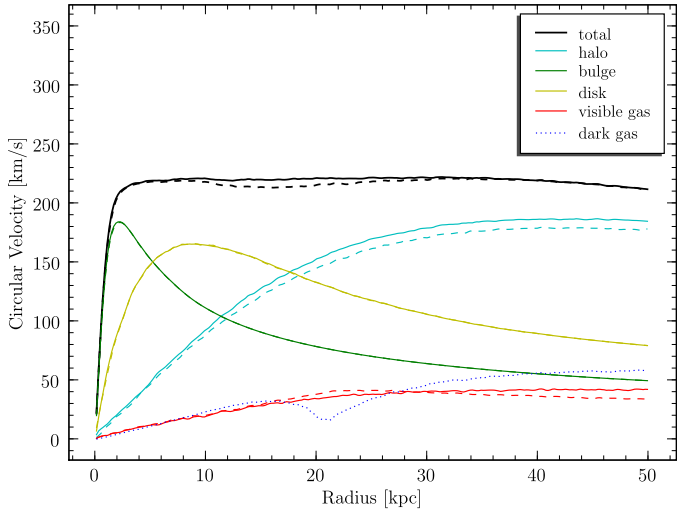}}
  \caption{Same figure as Fig.~\ref{crv_003} but for model $N=5$.}
  \label{crv_005}
  \end{figure}

Fig.~\ref{crv_003} and \ref{crv_005} compare the contribution of all
components to the rotation curve.  In the models with additional
baryons, the decreasing contribution of the dark halo mass in the
outer part is compensated by the dark gas, ensuring a flat rotation
curve up to $50\,\rm{kpc}$.  The bump in this component,
appearing below $25\,\rm{kpc}$, is due to the fact that, as no dark gas
resides in the central regions, the square velocity of this component
alone is negative. As in Fig.~\ref{crv_003} and \ref{crv_005} we have
plotted the absolute value of the velocity, the imaginary part appears
as positive.
Except around the transition radius of $20\,\rm{kpc}$, the total rotation
curve of models $N=3$ and $N=5$ remains nearly unchanged with respect
to the reference model (first rule).
\begin{table}
    \begin{tabular}{c c c c c c}
    \hline
    \hline
    N  &  $M_{\rm{vg}}$	         &  $M_{\rm{dg}}$              & $M_{\rm{b}}$ & $M_{\rm{Tot}}$             & $M_{\rm{b}}$/$M_{\rm{Tot}}$\\
       &  $[10^{11}\,\rm{M_\odot}]$  & $[10^{11}\,\rm{M_\odot}]$ &  $[10^{11}\,\rm{M_\odot}]$  & $[10^{11}\,\rm{M_\odot}]$ & \% \\
    \hline
    \hline
    1	    & 0.21	  & 0.00     & 1.18  & 6.36   &  18\\
    3	    & 0.12	  & 0.42     & 1.52  & 6.27   &  24\\
    5	    & 0.16	  & 0.66     & 1.80  & 6.12   &  29\\
    \hline
    \end{tabular}
    \caption[]{Parameters for the reference giant Sbc model.}
    \label{N135}
\end{table}

\subsubsection{$\Lambda$CDM substructures}\label{LCDM substructures}

In addition to the three previous models, we have built a fourth model
called the $N=1+\rm{s}$ model based on the $N=1$ models but taking
into account a sample of dark matter satellites predicted by the
$\Lambda$CDM cosmology, orbiting around the disk and interacting with
it.

The purpose of this model is to study the effect of $\Lambda$CDM
satellites on the spiral structure of the disk, and to compare it with
the effect of the additional baryons taken into account in models
$N=3$ and $N=5$.

\begin{figure}
  \resizebox{\hsize}{!}{\includegraphics[angle=0]{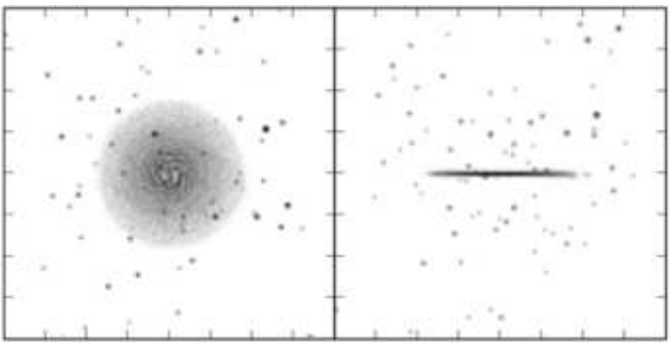}}
  \caption{Model $N=1+\rm{s}$ : 100 $\Lambda$CDM satellites around the
    gaseous disk of the $N=1$ model.  The images have been take after
    $t=2.2\,\rm{Gyr}$ of evolution. The box size is
    $400\times400\,\rm{kpc}$.}
  \label{satellites}
  \end{figure}

The $\Lambda$CDM satellites have been added following the technique
described by \citet{gauthier06} based on the distribution function of
\citet{gao04}, in agreement with recent $\Lambda$CDM simulations.
$100$ satellites extending up to $250\,\rm{kpc}$, with masses between
$1.5\times 10^{-4}$ and $0.02$ mass of the galaxy of the reference
model ($6.36\times 10^{11}\,\rm{M_\odot}$), have been added,
representing $10\,\rm{\%}$ of the total galactic mass.  Contrary to
\citet{gauthier06}, in our simulations, the substructures are Plummer
masses with a softening fixed to $0.25\,\rm{kpc}$.
Fig.~\ref{satellites} shows the surface density of model $N=1+\rm{s}$
including the $\Lambda$CDM satellites.  The darkness of the satellites
scales as a function of the mass.
%


\subsection{Velocity distribution}\label{velocity_distribution}

The velocity distributions are computed following the method proposed
by \citet{hernquist93}.  For the spherical distribution, the bulge,
the velocity dispersion is assumed to be isotropic and is derived from
the second moment of the Jeans equation \citep{binney87} which is
then:
\begin{equation}
         \sigma_{\rm{h}}^2(r) = \frac{1}{\rho_{\rm{h}}(r)}\int_r^\infty\! dr' \,\rho_{\rm{h}}(r')\, \partial_{r'} \Phi(r').
        \label{sr_sph}
\end{equation}
For the axisymmetric components (stellar disk and gas disk), we first
compute the vertical velocity dispersion $\sigma_z$, by satisfying the
Jeans equation in cylindrical coordinates \citep{binney87} separately
for each component $i$:
\begin{equation}
        {\sigma_z}_i^2 = \frac{1}{\rho_i(z)}\int_z^\infty\! dz' \,\rho_i(z')\, \partial_{z'} \Phi(z').       
        \label{sz_cyl}
\end{equation}
For the stellar disk and for the bulge, the radial velocity dispersion
$\sigma_R$ is fixed by imposing the value of the Safronov-Toomre
parameter $Q$ to $1$, ensuring the horizontal stability of the disk.
$Q$ is defined by:
\begin{equation}
        Q = \frac{\kappa\,\sigma_R}{3.36\, G\, (\Sigma_{d}  + \Sigma_{g})},
        \label{q_cyl}
\end{equation}
where $\kappa$ is the radial epicyclic frequency.  For the gas,
according the the HI observations, $\sigma_R$ is fixed to a value of
$10\,\rm{km/s}$.  The tangential velocity $\sigma_{\phi}$ is derived
from the epicycle approximation \citep[][ p.\,125]{binney87}:
\begin{equation}
        \frac{\sigma_{\phi}^2}{\sigma_R^2}
        = \frac{\kappa^2}{4\Omega^2}
        = \frac{R^{-3}\partial_R\left(R^3\partial_R \Phi\right)_{z=0}} 
             {4R^{-1}\left(\partial_R \Phi\right)_{z=0}},
        \label{sp_cyl}
\end{equation}
where the rotation frequency $\Omega$ is determined from the potential
$R$-derivatives as indicated in Eq.~(\ref{sp_cyl}).  Finally, the mean
azimuthal velocity is deduced from the second moment of the Jeans
equation:
\begin{equation}
        \langle v_{\phi}^2 \rangle = R\,\partial_{R} \Phi(R,z) + \sigma_R^2 - \sigma_\phi^2 + \frac{R}{\rho_i}  
	\partial_R \left( \rho_i \sigma_R^2  \right),
	\label{vm_cly}
\end{equation}
where we have assumed that the term $\langle v_R v_{\phi} \rangle = 0$.

\subsubsection{Velocity distribution for the visible and dark gas}

The radial velocity dispersions of the visible gas is set to be
constant at $10\,\rm{km/s}$ as observed:

\begin{equation}
        {\sigma_R}_{\rm{vg}} = 10\,\rm{km/s}.         
        \label{sR_vg}
\end{equation}
On the contrary, in order to ensure stability, the radial velocity
dispersions of the dark gas is set by imposing the Savronov-Toomre
parameter to be 1:
\begin{equation}
        {\sigma_R}_{\rm{dg}} = \frac{3.36\, G\, (\Sigma_{d} + \Sigma_{g}) \, Q}{\kappa}, \quad {\rm with} \quad Q=1.       
        \label{sR_wg}
\end{equation}
The velocity dispersions and the mean azimuthal velocity are then used
to distribute the model particles in the velocity space.


\section{Model evolution}\label{model_evolution}



The multiphase model has been implemented on the parallel Tree code
Gadget2 \citep{springel05}.  The models $N=1$, $N=1+\rm{s}$, $N=3$ and
$N=5$ contain respectively $319315$, $319415$, $535769$ and $731129$
particles.  The mass of the gas particles is constant between the
three models. The softening length is set to $250\,\rm{pc}$.  All
simulations have been run between $0$ and $4.5\,\rm{Gyr}$. For
simplicity, the feedback from supernovae has been turned off.

\subsection{Global properties}

Despite their different dark matter content, after $4.5\,\rm{Gyr}$,
the three models ($N=1$, $N=3$ and $N=5$) still share similar global
properties.

\begin{figure}
  \resizebox{\hsize}{!}{\includegraphics[angle=0]{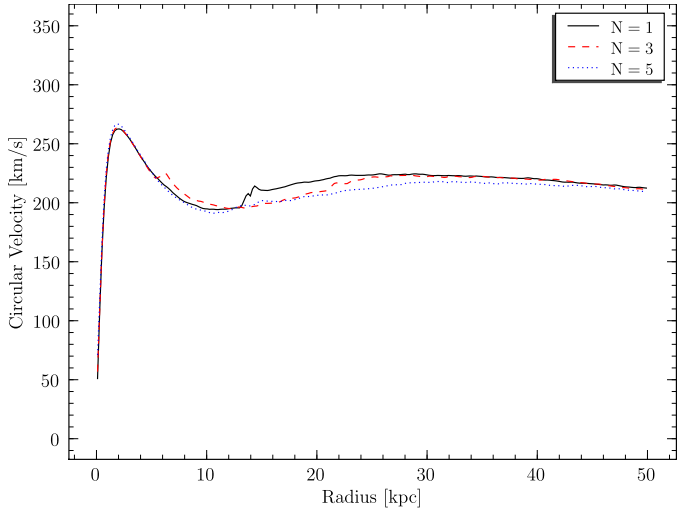}}
  \caption{Comparison of the rotation curve of the 3 models at $t=4.5\,\rm{Gyr}$.}
  \label{rotation_curves_0400}
  \end{figure}

\begin{figure}
  \resizebox{\hsize}{!}{\includegraphics[angle=0]{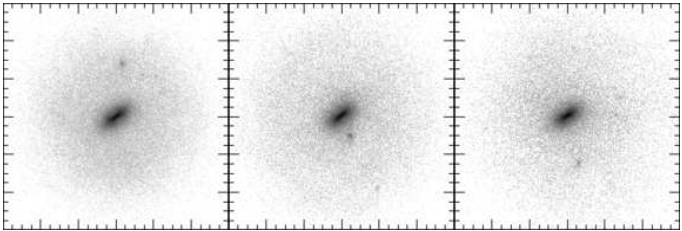}}
  \caption{Surface density map of the stellar disk of the 3 models ($N=1$, $N=3$ and $N=5$, from left to right) 
  at $t=4.5\,\rm{Gyr}$. The box size of each image is $60\times60\,\rm{kpc}$.}
  \label{stellar_bar}
  \end{figure}

In Fig.~\ref{rotation_curves_0400} we compare the total velocity curve 
of the three models, after $4.5\,\rm{Gyr}$ of evolution. In the
central part, they all show a bump corresponding to the presence of a
bar. In the outer part, all curves converge to the same values.  The
main differences occur around the transition radius ($\sim
10-20\,\rm{kpc}$), where models with additional baryons have slightly
lower values. However, these differences are simply a relic of the
differences existing in the initial conditions (see Fig.~\ref{crv_003}
and \ref{crv_005}).  The similarity of the rotation curves means that
the three models share the same radial potential dependency, having
similar horizontal epicyclic frequencies ($\kappa$ and $\Omega$).  The
pattern speed, the shape and the extension of the bar is also
identical (see Fig.~\ref{stellar_bar}).

\begin{figure}
  \resizebox{\hsize}{!}{\includegraphics[angle=0]{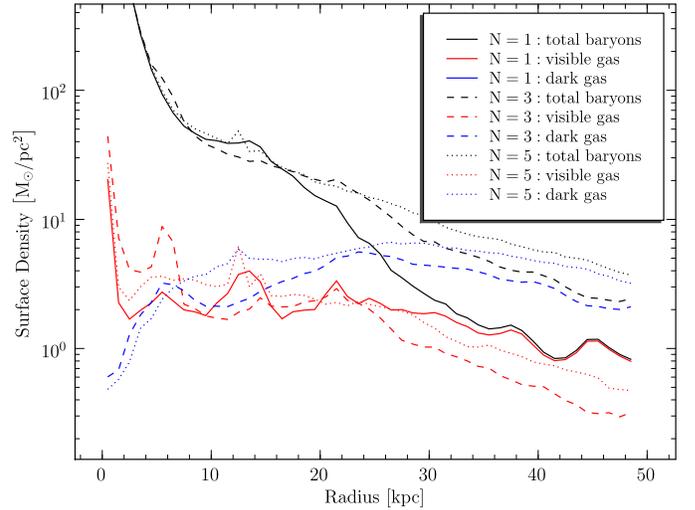}}
  \caption{Comparison of the surface density of different components at $t=4.5\,\rm{Gyr}$.}
  \label{surface_density_0400}
  \end{figure}

The surface density of the three models is compared in
Fig.~\ref{surface_density_0400}.  For radius smaller than
$30\,\rm{kpc}$, the curves corresponding to the visible gas surface
density are superimposed. Differentiating these three curves is only
possible in the far outer part, as was the case for the
initial conditions (see Fig.~\ref{den_003} and \ref{den_005}).
It is thus important to notice that the rotation curve and the
azimuthal averaged surface density are not sufficient to easily
distinguish between these three models, even after $4.5\,\rm{Gyr}$ of
evolution.

\begin{figure}
  \resizebox{\hsize}{!}{\includegraphics[angle=0]{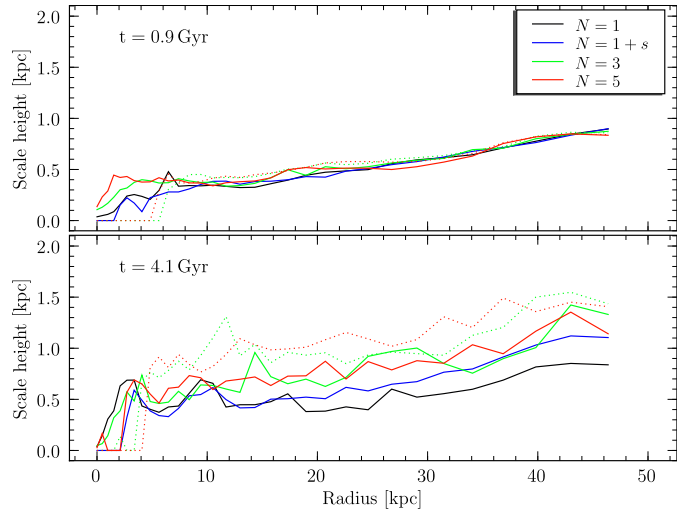}}
  \caption{Disk scale height as a function of the radius, for the four models at $t=0.9$ and $t=4.1\,\rm{Gyr}$. The lines
  corresponds to the visible gas while the dotted lines correspond to the dark gas.}
  \label{gplotcyl_r_hz}
  \end{figure}

In Fig.\ref{gplotcyl_r_hz} we compare the scale height of the disk of
the four models at $t=0.9$ and at $t=4.1\,\rm{Gyr}$.  The continuous
lines correspond to the visible gas while the dotted ones to the dark
gas. At $t=0.9\,\rm{Gyr}$, no difference exists between the four
models. After $4\,\rm{Gyr}$ of evolution the visible gas in the
$N=1+s$, $N=3$ and $N=5$ models has been slightly heated and presents higher
scale heights.  In the case of model $N=1+s$, the increase of the scale
height is attributed to the perturbation of the $\Lambda$CDM
satellites \citep{font01,dubinski08}, while for models $N=3$ and $N=5$
it is due to the coupling with the collisionless dark gas that has a
higher scale height (dotted lines).  From an observational point of
view, models $N=1+s$, $N=3$ and $N=5$ will be very similar.

\subsection{Disk stability and spiral structures}

\begin{figure}
  \resizebox{\hsize}{!}{\includegraphics[angle=0]{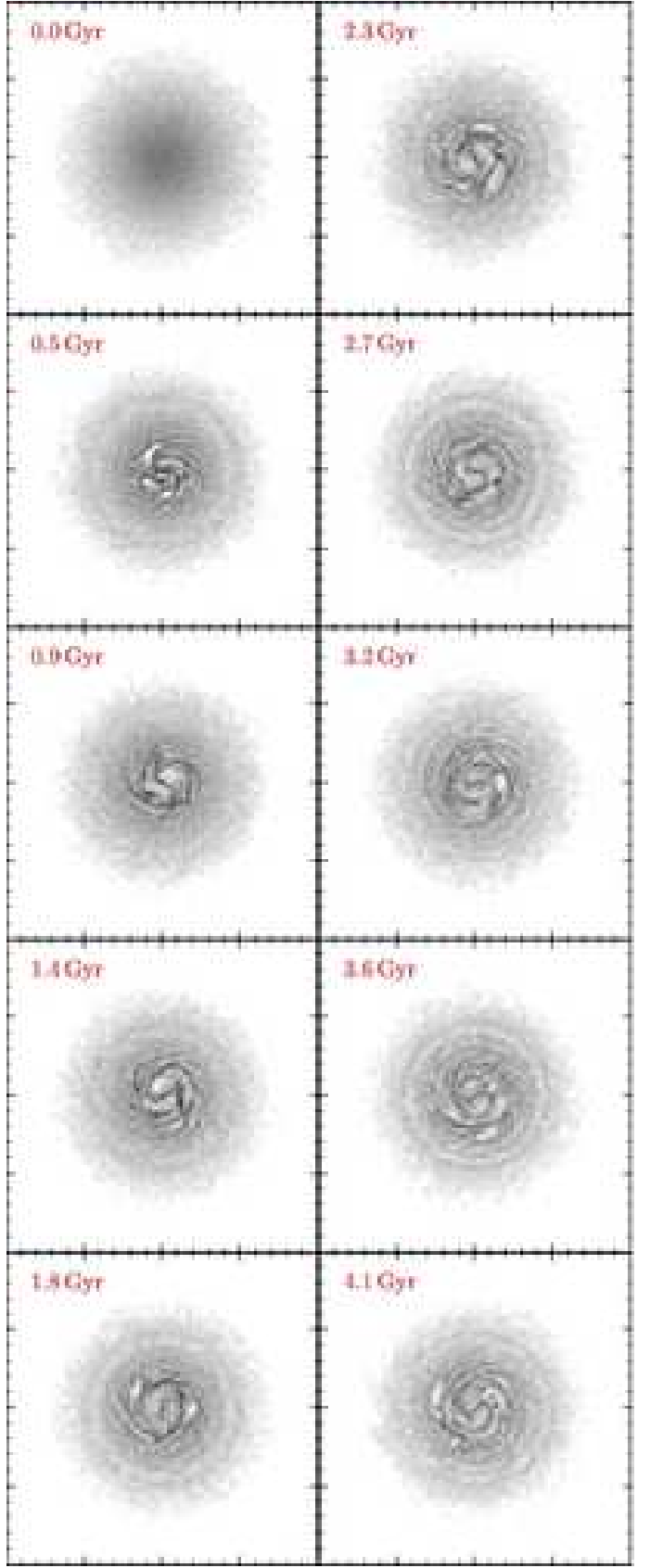}}
  \caption{Evolution of model $N=1$ between $0$ and
    $4.5\,\rm{Gyr}$. Each image represents the surface density of the
    visible gas. The box size is $200\times200\,\rm{kpc}$.}
  \label{evolution_001}
  \end{figure}

\begin{figure}
  \resizebox{\hsize}{!}{\includegraphics[angle=0]{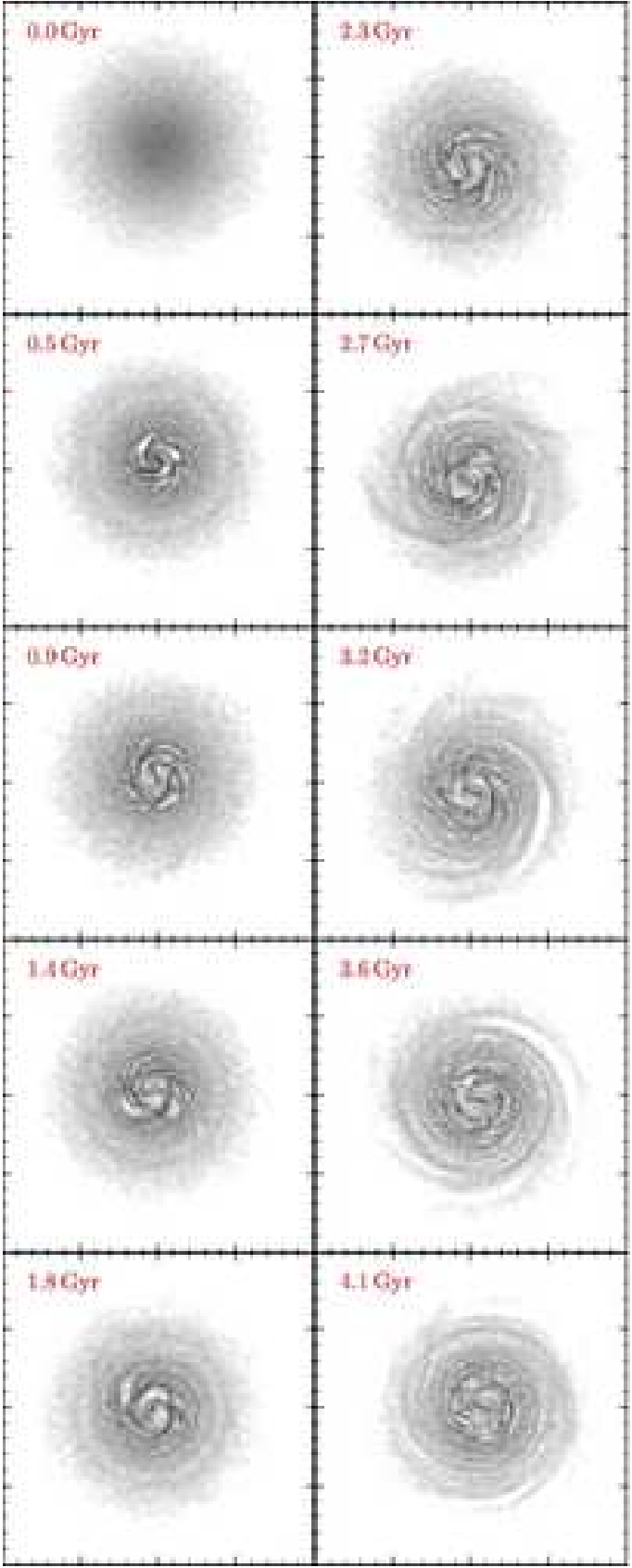}}
  \caption{Evolution of model $N=1$ including 100 $\Lambda$CDM
    substructures. Each image represents the surface density of the
    visible gas. The box size is $200\times200\,\rm{kpc}$.}
  \label{evolution_001s}
  \end{figure}

\begin{figure*}
  \resizebox{\hsize}{!}{\includegraphics[angle=0]{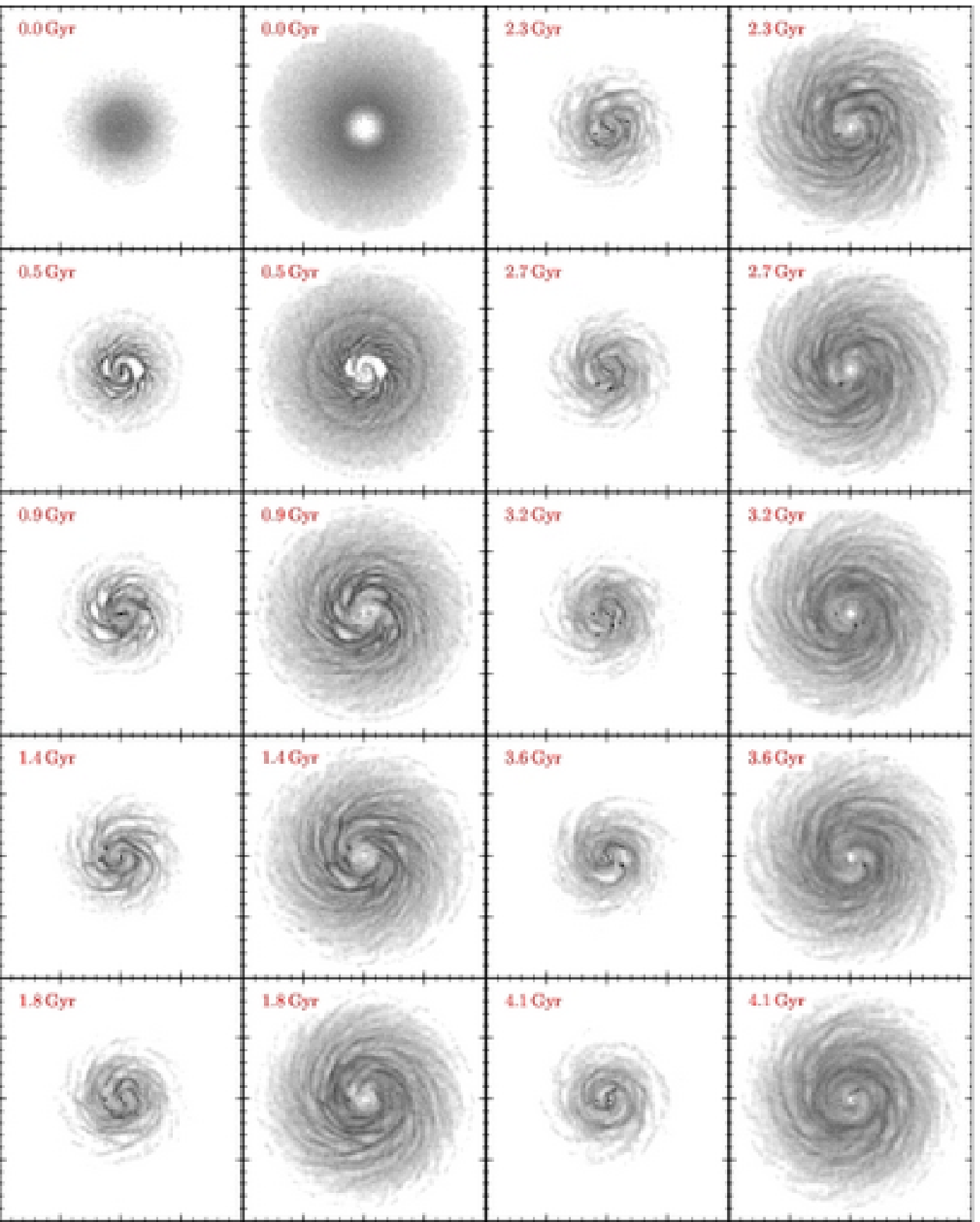}}
  \caption{Evolution of model $N=3$ between $0$ and $4.5\,\rm{Gyr}$.
    The first and third columns represent the visible gas surface
    density while the second and fourth represent the dark gas surface
    density.  The box size is $200\times200\,\rm{kpc}$.}
  \label{evolution_003}
  \end{figure*}

\begin{figure*}
  \resizebox{\hsize}{!}{\includegraphics[angle=0]{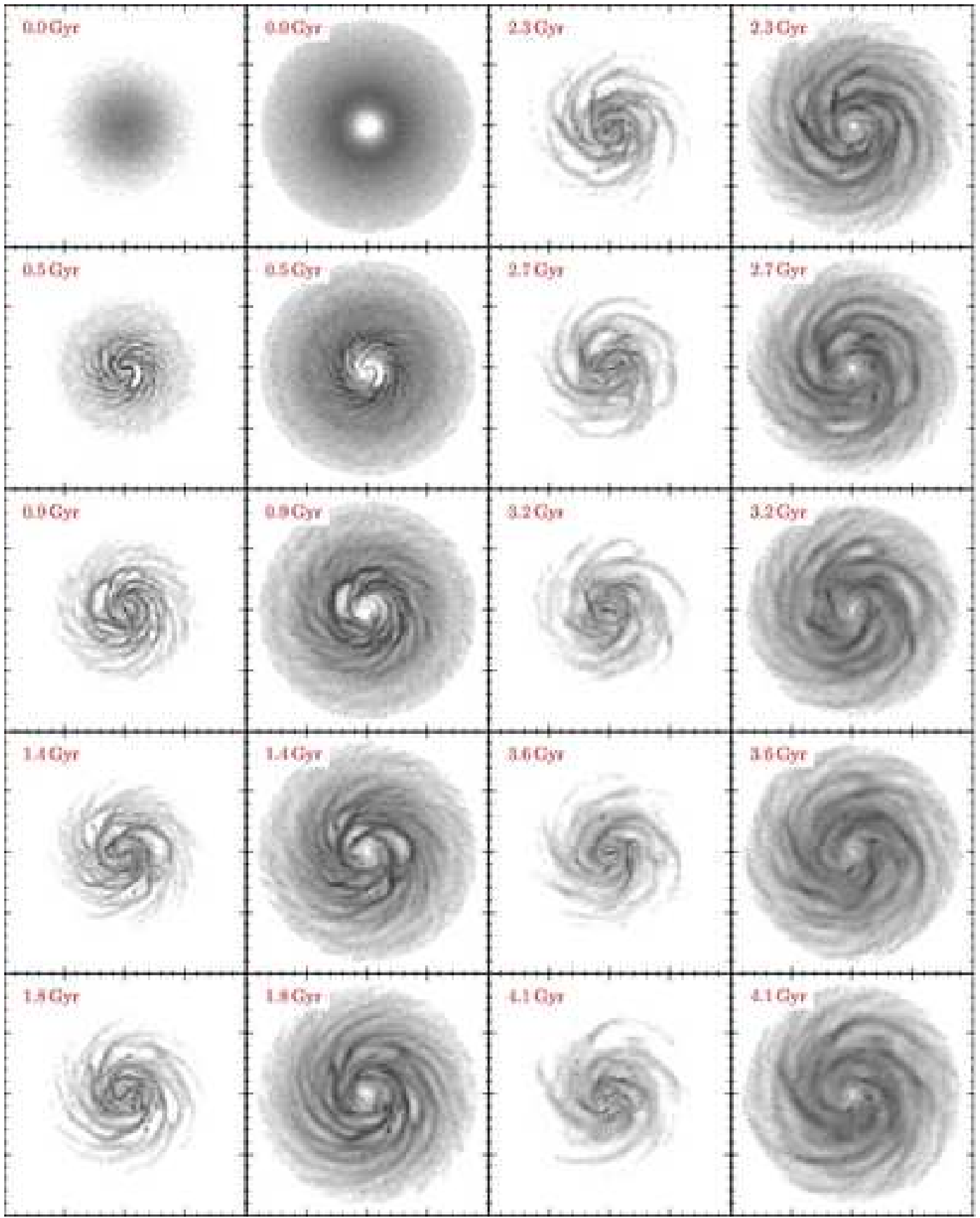}}
  \caption{Evolution of model $N=5$ between $0$ and $4.5\,\rm{Gyr}$.
    The first and third columns represent the visible gas surface
    density while the second and fourth represent the dark gas surface
    density.  The box size is $200\times200\,\rm{kpc}$.}
  \label{evolution_005}
  \end{figure*}

The main differences between the models with additional dark baryons
and the ``classical model'' appear when comparing the spiral structure
of the visible gas.  The surface density evolution of the three models
plus the one with the $\Lambda$CDM satellites is displayed in
Fig.~\ref{evolution_001}-\ref{evolution_005}.  In the two latter plots, in addition to the
visible gas surface density, the dark gas surface density is also
displayed.  These plots emphasize two important points. First, despite
the presence of larger amount of baryons in the outer parts, the disks
of models $N=3$ and $N=5$ are globally stable over the $4.5$ simulated
Gyrs.  Secondly, spiral structures have very different patterns and
may be used to distinguish between models with and without additional
baryons.

At all times during the evolution, the surface density of the dark gas decreases at the center, preserving a
hole. This hole simply means that no additional dark matter is
expected at the center, even after $4.5\,\rm{Gyr}$ of evolution.

\subsubsection{Disk stability}

\begin{figure}
  \resizebox{\hsize}{!}{\includegraphics[angle=0]{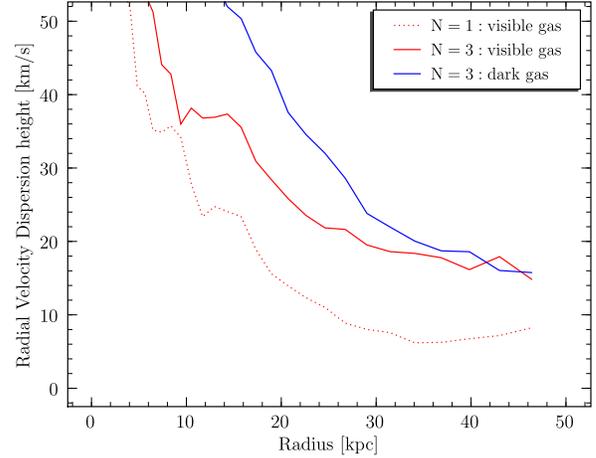}}
  \caption{Radial velocity dispersion at $t=4.5\,\rm{Gyr}$ of the
    visible (red) and dark (blue) gas components of model $N=3$.  The
    dotted line corresponds to the radial velocity dispersion of the
    visible gas of mode $N=1$.}
  \label{svr_003-102}
  \end{figure}

The stability of models $N=3$ and $N=5$ is ensured by the presence of
the two partially dynamically decoupled phases.  The dark phase has
higher velocity dispersions than the visible gas.  The radial velocity
dispersions of model $N=3$ are displayed in Fig.~\ref{svr_003-102}.
Because the dark gas is quasi non-collisional, its radial velocity
dispersion is much larger than the one of the visible gas of the
reference model (values dropping to $10\,\rm{km/s}$ in the outer
part.).  These higher values ensure the stability of the total
baryonic disk.  However, our multiphase model fails to reproduce the
low velocity dispersions expected in the visible gas.  While being
clearly decoupled from the dark component, the gas velocity dispersion
of model $N=3$ is higher than the one of the reference model,
especially in the outer part. This is the result of the weak coupling
due to the cycling between the dark and the visible gas.  The model
could be improved in the future, by simply increasing the stickiness
of particles in the outer part.

As discussed in Appendix~\ref{appendix2}, if one assumes that the dark
gas is as dissipative as the visible gas, the resulting velocity
dispersions of the total gas is no longer large enough to ensure the
disk stability.  In that case, the disk breaks and forms small clumps
(Fig.~\ref{evolution_000}).

\begin{figure}
  \resizebox{\hsize}{!}{\includegraphics[angle=0]{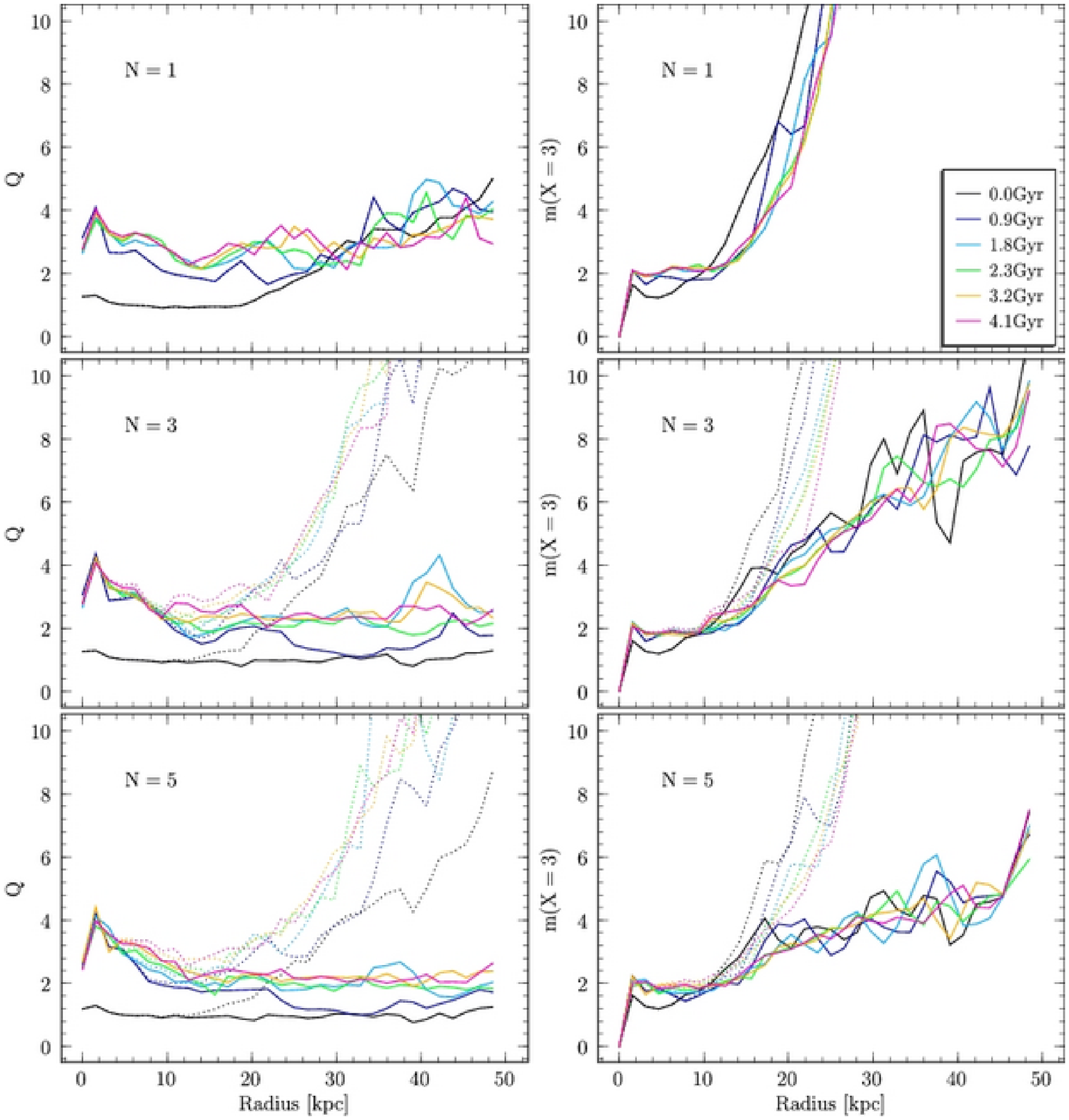}}
  \caption{ Stability of models $N=1$, $N=3$ and $N=5$ at different
    times ($0$, $0.9$, $1.8$, $2.3$, $3.2$ and $4.1\,\rm{Gyr}$).  The
    left column show the Toomre parameter (Eq.~\ref{Qeff}).  The right
    column shows the mode corresponding to the swing amplification
    parameter $X=3$. Modes higher than the latter may be amplified
    while lower modes are stable.  In both columns, the dotted lines
    correspond to the values computed without taking into account the
    dark gas.
  The corresponding color code for all times are given in the upper right panel.
}
  \label{XQ}
  \end{figure}

Since our disk model is made of three dynamically decoupled
components (the stellar disk, the visible disk and the dark disk)
having distinct velocity dispersions, a precise study of its stability
would require to use a multi-component stability criterion.  The
stability of multi-component galactic disks has been discussed by
\citet{jog84a,jog84b} and the equivalent of the Savronov-Toomre
parameter may be computed \citep{jog96}. While this criterion is
theoretically valid for an $n$-components system the computational
procedure described by \citet{jog96} is only valid for a
two-components system (See also \citet{wang94} and
\citet{elmegreen95}).  We have instead used a Savronov-Toomre
($Q_{\rm{1-c}}$) assuming a single component.  Formally, this
$Q_{\rm{1-c}}$ is defined by:
\begin{equation}
        Q_{\rm{1-c}}(R) \simeq \frac{\kappa\, \sigma_{R,\rm{disk}}}{3.36\,G\,\Sigma_{\rm{disk}}},       
        \label{Qeff}
\end{equation}
where $\Sigma_{\rm{disk}}$ and $\sigma_{R,\rm{disk}}$ are computed
over the visible gas, dark gas and stars particles, while $\kappa$
results from the total potential (including halo dark matter and disk
dark baryons).  The left panels of Fig.~\ref{XQ} shows the parameter
$Q_{\rm{1-c}}$ for the four models at different times.  It is clear
that $Q_{\rm{1-c}}$ is always larger than $1$, explaining the
stability observed in Fig.~\ref{evolution_003} and
\ref{evolution_005}.
%

%

\subsection{Spiral structure and swing amplification}


While being larger than $1$, the two models with additional baryons
are characterized by a nearly constant $Q_{\rm{1-c}}$ from $10$ to
$50\,\rm{kpc}$, with values between $2$ and $3$ for model $N=3$ and
around $2$ for model $N=5$.  With these rather low values, the disks
are only quasi-stable. Contrasting and open spiral structures are
continuously generated up to the end of the dark gas disk at
$100\,\rm{kpc}$.  These spirals are thus naturally present in the
visible gas up to its end, where its surface density drops, in
agreement with most HI observed disks. See for example the
impressive case of NGC\,6946 \citep{boomsma08}.

On the contrary, model $N=1$ has a higher $Q_{\rm{1-c}}$, increasing
well above $3$ for $R>30\,\rm{kpc}$.  Consequently, the outer disk is
more stable, preventing the formation of spirals, as observed in
Fig.~\ref{evolution_001}.  This difference is the key point that
allow us to distinguish between models with and without additional
baryons.

\begin{figure}
  \resizebox{\hsize}{!}{\includegraphics[angle=0]{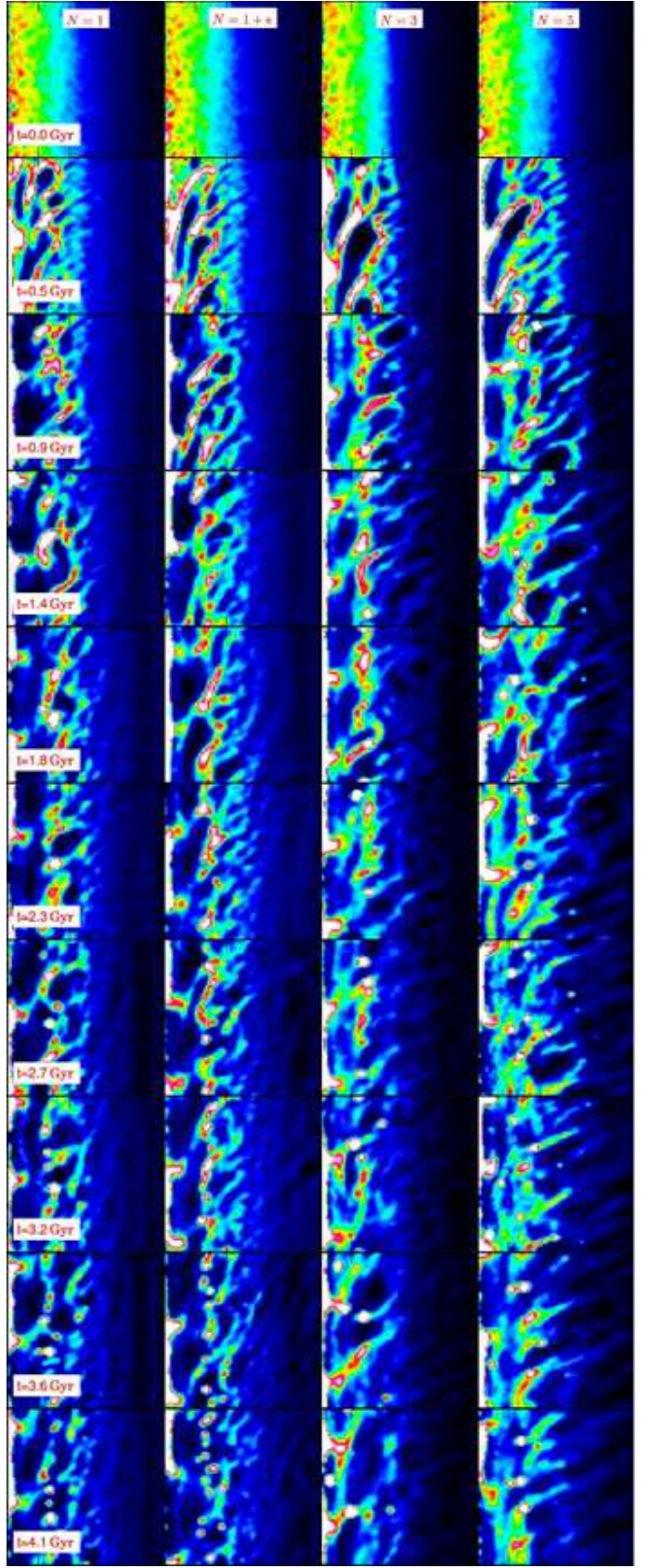}}
  \caption{Evolution of the spiral structure in a ($R-\phi)$-plot of the four models. 
  Each image represents the surface density of the visible gas. The abcissa corresponds to the radius $R$ (from $0$ to $50\,\rm{kpc}$) while
  the ordinate to the azimuthal angle (from $0$ to $2\pi$).}
  \label{spiralt}
  \end{figure}
\begin{figure}
  \resizebox{\hsize}{!}{\includegraphics[angle=0]{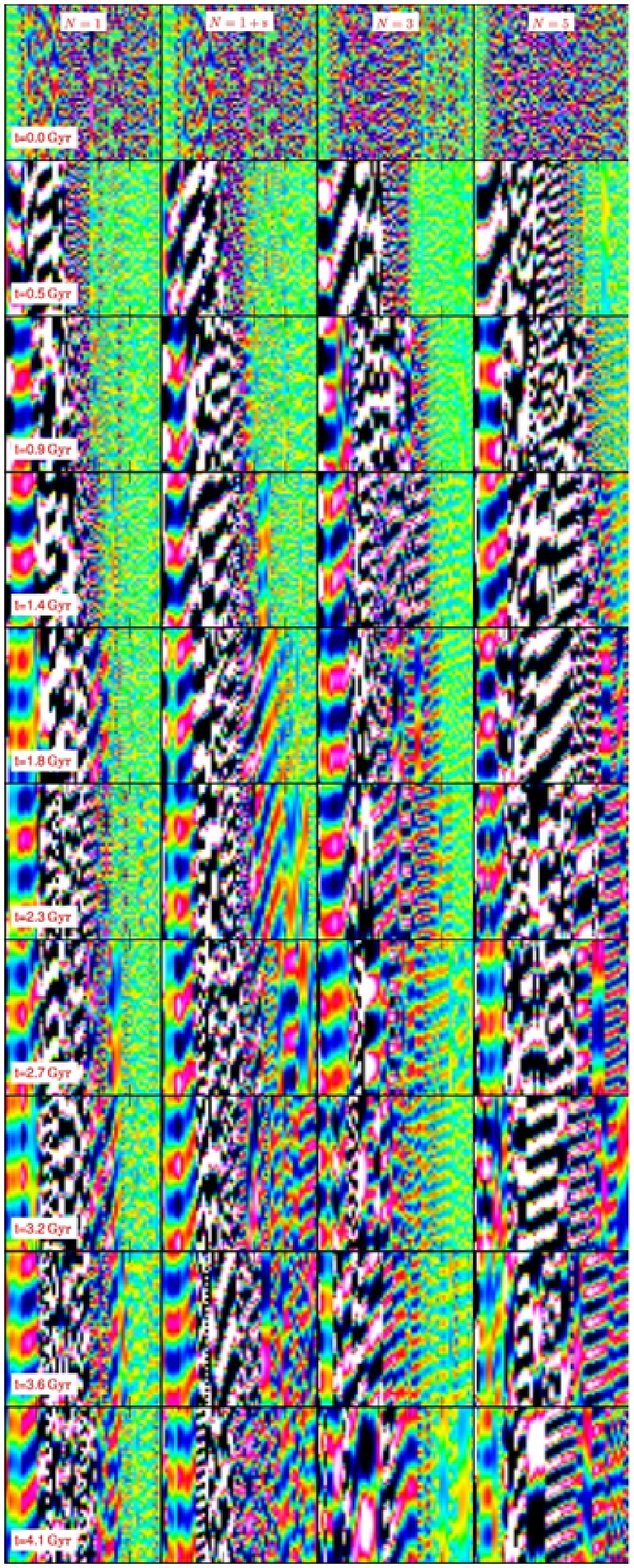}}
  \caption{Evolution of the spiral structure in a ($R-\phi)$-plot of the four models. 
  Each image represent the dominant mode of Fig.\,\ref{spiralt}.}
  \label{spiralm}
  \end{figure}

In order to improve our understanding of the spiral structure, we have
computed polar maps (($R-\phi)$-plot) of the visible gas surface
density (from $R=0$ to $R=50\,\rm{kpc}$) at different times
(Fig.~\ref{spiralt}).  These maps can be, for example, compared to
the HI observation in M\,83 (Fig.~$12$ of \citet{crosthwaite02}).
We then performed a Fourier decomposition of those maps, for each
radius.  The spiral structure can be represented by each dominant
azimuthal modes for a given radius, i.e. the mode with the largest
amplitude, excluding the $m=0$ mode.  Fig.~\ref{spiralm} displays the
amplitude of the dominant azimuthal modes found for each plot of
Fig.~\ref{spiralt}.  For a direct comparison between the models, in
all plots, the signal amplitude is coded with the same colors.
In polar plots, a regular spiral structure of mode $m$ appears as
$2\,m$ inclined parallel lines\footnote{Here, a line is either a
  maximum or a minimum.}, where the inclination (that may vary with
radius) gives the pitch angle $i$ of the arm ($i = \pi/2 -
\rm{arctan}(p)$, where $p$ is the slope of the lines).  Such regular
features are present for the $N=3$ and $N=5$ models, indicating the
presence of rather open spiral arms (large $i$) that may be followed
up to $50\,\rm{kpc}$ at nearly all times.  The number of arms is not
constant and oscillates between $3$ and $8$ (model $N=3$) and $3$ and
$6$ (model $N=5$), usually increasing with increasing radius
\citep{bottema03}.  Such large features are not observed for the $N=1$
model. The spiral arms disappear around $R=25\,\rm{kpc}$.

Spiral arms in a disk may be the result of `swing
amplification'\citep{goldreich65,julian66,toomre81}, a mechanism that
locally enhances the self-gravity and leads to the amplification of a
small perturbation in a differentially rotating structure.  The
amplification results from the resonance between the epicyclic motions
of stars (or gas) and the rate of change of the pitch angle of a
density wave, during its conversion from leading to trailing
\citep{binney87}.
The swing amplification mechanism may be quantified by the parameter
$X$, the ratio of the perturbation wavelength and a critical
wavelength:
\begin{equation}
        X = \lambda/\lambda_{\rm{c}}, \qquad \textrm{where} \qquad \lambda_{\rm{c}} = \frac{4\,\pi^2\,G\,\Sigma}{\kappa^2}.        
        \label{Xlambda}
\end{equation}
Since the wavelength $\lambda$ of a spiral pattern of mode $m$ in a disk is $2\,\pi\,R/m$, $X$ becomes:
\begin{equation}
        X(R) = \frac{R\,\kappa^2}{2\,\pi\,G\,\Sigma\,m}.       
        \label{X}
\end{equation}
%
Similarly to the $Q_{\rm{1-c}}$ computation, we have computed the
swing amplification parameter $X$, taking into account only the
surface density of the disk ($\Sigma = \Sigma_{\rm{disk}}$).  As it is
too dynamically hot, at first order, the dark matter is not sensitive
to the density perturbation and its effect is not included in the
computation of $X$.  If we assume that the swing amplification is
efficient for $X \lesssim 3$ \citep[Fig.~7]{toomre81,athanassoula84}
\footnote{In fact the exact critical value may vary, depending on the
  parameter $Q_{\rm{1-c}}$ and on the exact velocity curve through the
  shear parameter $\Gamma=-\frac{r}{\Omega}\frac{d}{dr}\Omega$ which
  in our case is nearly equal to $1$.}, we can reverse Eq.~\ref{X} and
compute the value of the critical mode $m(X=3)$:
\begin{equation}
        m(X=3) = \frac{R\,\kappa^2}{6\,\pi\,G\,\Sigma_{\rm{disk}}}.       
        \label{mX}
\end{equation}
The meaning of the critical mode is the following: modes higher than
$m(X=3)$ may be amplified by the swing amplification mechanism.  On the contrary,
the ones above $m(X=3)$ are stable and thus no spiral pattern with
$m<m(X=3)$ is observed.

The right panels of Fig.~\ref{XQ} displays the critical modes at
different times.  For the $N=1$ model, $m(X=3)$ increases very
quickly, being above $6$ for $R>20\,\rm{kpc}$.  At larger radii, only
large $m$ modes, corresponding to noise, may be amplified by the
self-gravity of the disk.  No spiral pattern is expected, in agreement
with Fig.~\ref{spiralm} and \ref{evolution_001}.

When increasing $\Sigma_{\rm{disk}}$ by including additional baryons,
self-gravity in increased and lower modes may be amplified at larger
radius.  For example, $m=6$ modes may still be amplified up to
$R=35\,\rm{kpc}$ in model $N=3$ and up to $R=50\,\rm{kpc}$ in model
$N=5$.  This explains the patterns present in
Fig.~\ref{spiralm} and the spiral structure of
Fig.~\ref{evolution_003} and Fig.~\ref{evolution_005}.

The dotted lines of Fig.~\ref{XQ} (both in the left and right panels)
correspond to the $Q_{\rm{1-c}}$ and $m(X=3)$ values as would be
deduced by an observer ignoring the presence of the dark gas and
taking into account only the visible gas and stars in Eq.~\ref{Qeff}
and \ref{mX}.  Clearly, $Q_{\rm{1-c}}$ is well above $4$ for a radius
larger than $20\,\rm{kpc}$, indicating that the disk should be stable.
The $m(X=3)$ parameter implies that modes with low
perturbation should be prohibited by the swing amplification theory,
in contradiction to the observed spiral structure.  This point may
explain why spiral structures are present in galaxies where the
$Q_{\rm{1-c}}$ (or $X$) parameter deduced from the visible component
is well above $5$ (resp. 3), as is the case, for example, for the
galaxy NGC\,2915 \citep{bureau99}.
 

\subsubsection{The effect of the $\Lambda$CDM satellites}

We can now ask if the perturbations generated by the $\Lambda$CDM
satellites are able to reproduce the large scale spiral patterns of
galaxies, comparable with the ones obtained when including additional
baryons.

The effect of the satellites may be seen by comparing the surface
density of the visible gas disk of model $N=1$
(Fig.~\ref{evolution_001}) and model $N=1+s$
(Fig.~\ref{evolution_001s}).  Perturbations in the far outer part of
the disk are visible between $2.3$ and $4.1\,\rm{Gyr}$ and appear as
asymmetric and winding spiral arms.
These perturbations are better traced in Fig.~\ref{spiralm}, where the
same Fourier analysis has been performed for model $N=1+s$.  There,
the effect of the satellite is already perceptible at
$t=1.4\,\rm{Gyr}$ and $R\sim 30\,\rm{kpc}$.  Between $t=1.4$ and
$t=2.3\,\rm{Gyr}$ an $m=2$ mode develops and disappears
afterwards.  The large slope of the modes indicates a very small pitch
angle.  Another $m=2$ mode appears briefly between $t=2.3$ and
$t=2.7\,\rm{Gyr}$ ($R > 40\,\rm{kpc}$).  In this plot, no clear
dominant mode is seen afterwards, between $25$ and $50\,\rm{kpc}$.
However, as the Fourier decomposition was performed within
$R=50\,\rm{kpc}$, we are missing the spirals observed at larger radii
in Fig.~\ref{evolution_001s} at $t=3.6\,\rm{kpc}$.

The comparison between the spirals resulting from the self-gravity of
the disk or the ones formed by satellites perturbations reveals two
important differences:

\begin{enumerate}

\item As it does not result from a gravity wave, a spiral arm
  generated by the perturbation due to a satellite is affected by the
  differential rotation of the disk. Consequently the arm winds
  quickly (the pitch angle is small) and disappears in a dynamical
  time (winding problem, see for example \citet{binney87}).

\item Spiral patterns generated by the swing amplification are
  globally continuous along the disk.  We have seen for example that
  the dominant modes slowly decrease outwards.  As they result from a
  local perturbation, asymmetries and strong discontinuities exist in
  the pattern of spirals induced by satellites. An example is seen in
  Fig.~\ref{spiralm} at $t=2.7\,\rm{Gyr}$.  No dominant mode exists
  between $25$ and $40\,\rm{kpc}$ while an $m=2$ mode is present at
  larger radius, disconnected from the rest of the disk.

\end{enumerate}

In conclusion, the $\Lambda$CDM satellites are able to perturb a disk
and induce spiral arms. However, these arms wind up and disapear
quickly. Even if they are periodically induced by satellites, they are
unable to reproduce a regular and extended spiral patterns.

According to the theory of spiral structure formation, the fact
that $\Lambda$CDM induces only localised and decorrelated
perturbations is not surprising.  Indeed, if satellites act as
triggers for disk instabilities, the subsequent amplification of the
instabilities leading to the formation of a large scale pattern is
determined by the property of the disk itself: its self-gravity
(surface density), its velocity curve and its velocity dispersions,
the three quantities involved in the $Q_{\rm{1-c}}$ and $X$
computation.  Without the interplay of these essential ingredients,
only discontinuous local structures, resulting from the random impact
of the satellites on the disk, will be observed.


\section{Discussion and conclusion}\label{conclusions}


%
%
%
%
%

Using a new self-consistent N-body multiphase model, we have shown
that the hypothesis where galaxies are assumed to have an additional
dark baryonic component is in agreement with spiral galaxy properties.
The key element of the model is to assume that the ISM is composed out
of two partially dynamically decoupled phases, the observed
dissipative gas phase and a dark, very cold, clumpy and weakly
collisional phase.  
Motivations for invoking a similar component made of cold invisible gas, 
like the disk-halo conspiracy, the HI-dark matter proportionality,
extreme scattering and microlensing events, the hydrostatic equilibrium of the Galaxy
or the survival of small molecular structures without shielding,
have been given many times in the literature
\citep[e.g.,][]{pfenniger94b,henriksen95,gerhard96,walker98,
kalberla98,kerins02,heithausen02,heithausen04}.

An original scheme, designed to overcome numerical limitations,
has been proposed to compute the cycling between these two phases when
subject to heating and cooling.  From this scheme, we have shown that
realistic self-consistent N-body models of spiral galaxies containing
an additional baryonic dark matter content may be constructed.  These
models share similar observational properties with classical CDM
disks, like the rotation curve and visible gas surface density and in
that sense, are observationally similar.

The main result of this work is that, despite having more mass in the
disk, these systems are globally stable, the stability being ensured
by the larger velocity dispersion of the dark gas that dominates the
gravity in the outer part of the disk.  In addition, the enhanced
self-gravity of these disks, due to the presence of the dark clumpy
gas, makes them more prone to form spirals extending up to
$100\,\rm{kpc}$ in the dark gas.  The spiral structure is revealed by
the HI that acts as a tracer of the dark gas, up to the radius where
its surface density becomes so small that the hydrogen is fully
ionized and hardly observable.  This gives a natural solution of the
numerous observations of HI unstable disks that are difficult to
explain by the self-gravity of the HI disk alone (see for example
\citet{bureau99,masset03}).

Depending on the theory of spiral structure (the local stability of
differentially rotating disks and the swing amplification mechanism) our
results depend on the three quantities involved in the Toomre
parameters: the disk self-gravity by its surface density, its rotation
curve and its radial velocity dispersion.  Consequently, these results
will not be affected by a different parametrisation of the stellar
disk or the dark halo, as long as a similar rotation curve is
reproduced.

We have seen that in our models the velocity dispersion of the
collisionless or weakly collisional components increases with time,
similarly to what stars are known to do in the Milky Way in the so-called
Wielen's diffusion.  This is fully expected from our 
general understanding of the evolution of self-gravitating disks of
collisionless particles, where the effects of spiral arms and
especially a bar are sufficient to cause such a radial heating. 
A more detailed discussion and references 
are given in \citet{sellwood02}.

While the cold gas phase is probably weakly collisional, we have treated it as strictly
collisionless. How will the spiral morphology then be modified by introducing a weak 
dissipation in this component ?  In Appendix~\ref{appendix2} we present the case where
the dark component is as dissipative as the observed gas. In this extreme case, the disk is 
strongly unstable and fragments as a consequence of the Jeans instability.  At some point, 
decreasing the dissipation of the dark gas strengthens the global stability, while
reinforcing the spiral arm contrast. 
The dynamical coupling of the two phases through the cycling is as important as the dissipation.
A very short transition timescale $\tau$ leads to a strong
coupling.  Additional simulations show that in that case, the velocity dispersions of
the two components are  nearly similar, the one of the observed gas being higher.
Consequently, the spiral structures are less contrasting. On the contrary, if the two phases
are decoupled, the observed gas velocity dispersion is smaller, and the arm contrast
higher. In both cases however, the large scale spiral pattern remains similar.

Additional simulations have investigated the effect of $\Lambda$CDM
substructures on the outer HI disk.  Satellites only generate local,
winding and short-lived arms.  Because the HI disk is not sufficiently
self-gravitating, its response is too small to amplify the
perturbations.  In that sense, we confirm the results of
\citet{dubinski08}. The tidal effects of the satellites are generally
small and so are not responsible for large scale spiral patterns.

If models of galactic disks with additional dark baryons are in agreement with
observations, we must ask whether they are in agreement with the
$\Lambda$CDM scenario.  From the cosmic baryon budget
\citep{fukugita04}, we know that nearly $60\%$ of the baryons
predicted from primordial nucleosynthesis are not observed.
Multiplying the galactic baryon content (about $8\%$) by a factor of
$2$, as it is assumed in our most massive $N=5$ model, will still be
in agreement with the baryon budget.  From a galactic point of view,
multiplying the baryon content by a factor 2  is well inside the
observed uncertainties.  For a circular velocity curve of about
$250\,\rm{km/s}$ typical of our models, the baryon content varies by
more than a factor of $5$ (see Fig.~$1$ of \citet{mayer04}). Assuming
a corresponding virial mass of $10^{12}\,\rm{M_\odot}$ the cosmic
baryon fraction may vary around $10$ and $50\%$. With a baryonic
fraction of $29\%$, our $N=5$ model corresponds to a large but
realistic value.

Numerical simulations of the formation of large scale structures
predict that the ``missing baryons" reside in a warm-hot gas phase in
the over-dense cosmic filaments \citep{cen99,cen06}. However, there are
now also signs of accretion of cold gas during the build up of
galactic disks \citep{keres05}.  As discussed in
Section~\ref{multiphase_model}, a low density resolution leads to an
under-estimate of the gas cooling, missing over-dense regions
present in an inhomogeneous ISM, where the cooling time of the gas is
very short.  Since the assumed amount of dark baryons is not in
contradiction with the cosmic baryon budget, we need to wait for
future works to see if the correct treatment of inhomogeneous ISM,
including shocks, is able to cool $10\%$ of the ``missing baryons"
during the hierarchical structure formation, making the present model
consistent with the $\Lambda$CDM scenario.

In a companion paper, we have shown that our model also explains
the puzzling presence of dark matter in the collisional debris from
galaxies \citep{bournaud07}.

The remaining questions are obviously why the cold dark phase
is invisible and what the physics ruling it might consist of.  
Since star formation is also a poorly understood process, but we know that stars do form
from the coldest observable molecular gas, the question of why stars are
not supposed to form in the coldest form of our invoked dark gas
remains open.  Suggestions of why star formation may not always
start in cold gas have been given in \citet{pfenniger94b}.
Essentially, if molecular gas fragmentation goes down to sub-stellar
mass clumps staying cold, these clumps are unable to free nuclear
energy and become stars. 
Since they are close to the $3\,\rm{K}$ background, they almost do not radiate.
Because of their negative specific capacity,
such self-gravitating clumps do not increase their temperature when
subject to external heating, but stay cold while evaporating, a
sometimes overlooked feature of self-gravity.  Many earlier works have
discussed these issues, for example how cold molecular hydrogem may
clump in dense structures of solar system size but stay undetected
\citep{combes97,pfenniger94b,pfenniger04}.  However, much more work
remains to understand the precise physics of very cold gas ($ T <
10\,\rm{K}$) in low excitation regions, like the outer galactic disks, but
also in planetary nebulae and star forming regions where observed
dense cold gas clumps self-shield from outer radiation and reach
sub-$10\,\rm{K}$ temperatures.  It is generally known that H$_2$ condenses in
solid form even at interstellar pressures at temperatures close to the
$3\,\rm{K}$ cosmic background.  A phase transition means that a richer
physics must be expected. But since H$_2$ is usually mixed with about
10\% of helium in mass, it was for many years not clear how to
describe the equation of state of this mixture in such conditions.
Recently, \citet{safa08} succeeded in describing this mixture for astrophysically
interesting conditions with chemo-physical methods, reproducing its
main characteristics like the critical point and the condensation
curve, as well as predicting the conditions of He-H$_2$ separation.
Work is in progress to apply these results 
to the cold interstellar gas.


\begin{acknowledgements}
  It is a pleasure to thank Chandra Jog for discussions on the
  stability of multi-components systems and Jerry Sellwood for
  motivating us to perform Fourier analysis.  
  The simulations have been run on the Horizon mini grid and the Horizon
  meso-machine in Paris, as well as on the Regor cluster at the Geneva Observatory.
  Data reduction and galaxy maps have been performed using the parallelized Python
  pNbody package (see http://obswww.unige.ch/~revaz/pNbody/).
  This work was supported by the Swiss National Science Foundation and by the French
  Centre National Pour la Recherche Scientifique (CNRS).
\end{acknowledgements}


\begin{appendix}


\section{The Miyamoto-Nagai potential : another parametrization}\label{appendix1}

The Miyamoto-Nagai potential \citep{miyamoto75} in cylindrical
coordinates is usually written as:

\begin{equation}
  \Phi^{\rm{MN}}(R,z) = - \frac{GM}{\sqrt{R^2 + \left(a+\sqrt{z^2+b^2}\right)^2 }},
  \label{mn_pot}
\end{equation}
where $M$ is the total mass and $a$ and $b$ are two parameters.  It is
useful to parametrize the Miyamoto-Nagai potential, introducing the
horizontal scale length $h_R$ and the vertical scale length $h_z$
defined by:
\begin{equation}
   h_R=a+b \quad \textrm{and} \quad h_z=b.
  \label{mn_ab}
\end{equation}
The potential then becomes:
\begin{equation}
  \Phi^{\rm{MN}}_{(M,h_z,h_R)}(R,z) = - \frac{GM}{\sqrt{R^2 + \left(h_R-h_z+\sqrt{z^2+h_z^2}\right)^2 }}.
  \label{mn_pot_new}
\end{equation}


%
Using this parametrization, the circular velocity curve is written as:
\begin{equation}
  v_{\rm{c}}^2 =  R \frac{\partial}{\partial R}\Phi^{\rm{MN}}(R,0) =\frac{GMR^2}{\left[ R^2 + h_R^2 \right]^{3/2}},
  \label{mn_v2}
\end{equation}
which is independent of the vertical scale length $h_z$.  This means
that any Miyamoto-Nagai model with mass $M$ and horizontal scale
length $h_R$ will fit the same rotation curve. One may also recognize
in Eq.~\ref{mn_v2} the familiar velocity curve of the Plummer model,
the latter being a subclass of the Miyamoto-Nagai model, with $a=0$
and $b=h_R$).

A corollary of Eq.~\ref{mn_v2} is that the rotation curve of any
Miyamoto-Nagai model can be reproduced by a linear combination of two
or more Miyamoto-Nagai models, having different scale heights and
appropriate masses. For example, the potential $\Phi'_{M}(R,z)$
defined by~
\begin{equation}
  \Phi'_{M}(R,z) = f \Phi^{\rm{MN}}_{(M,h_z',h_R)}(R,z) + (1-f) \Phi^{\rm{MN}}_{(M,h_z,h_R)}(R,z),
  \label{mn_pot_sum}
\end{equation}
where $f\in[0,1]$ and $h_z'\ne h_z$, will have the same velocity curve as $\Phi^{\rm{MN}}_{(M,h_z,h_R)}(R,z)$.

\section{Evolution of model $N=3'$ }\label{appendix2}

In this appendix we present the evolution of model $N=3'$, where all
the gas (visible and dark) is assumed to be dissipative, computed with
the sticky particle scheme (see Section~\ref{visible_and_dark}).  The
sticky parameters used are similar to the ones used for the visible gas
in model $N=3$ and $N=5$.  The evolution shows that at the beginning
($t=0-t=0.8\,\rm{Gyr}$) the strongest dissipation reinforces the
spiral constant in the central part of the galaxy. But after
$1\,\rm{Gyr}$ the inner disk, which is dynamically very cold, becomes
strongly unstable. The spirals break into several dense clumps and the
disk does not look like a real galaxy. This effect is even stronger
for an $N=5$ model, which is not displayed here.  This test shows that
the dark matter in the disk cannot be as dissipative as the visible
disk.  Otherwise, the disk is unstable and fragments in less than a
dynamical time.
  \begin{figure}
  \resizebox{\hsize}{!}{\includegraphics[angle=0]{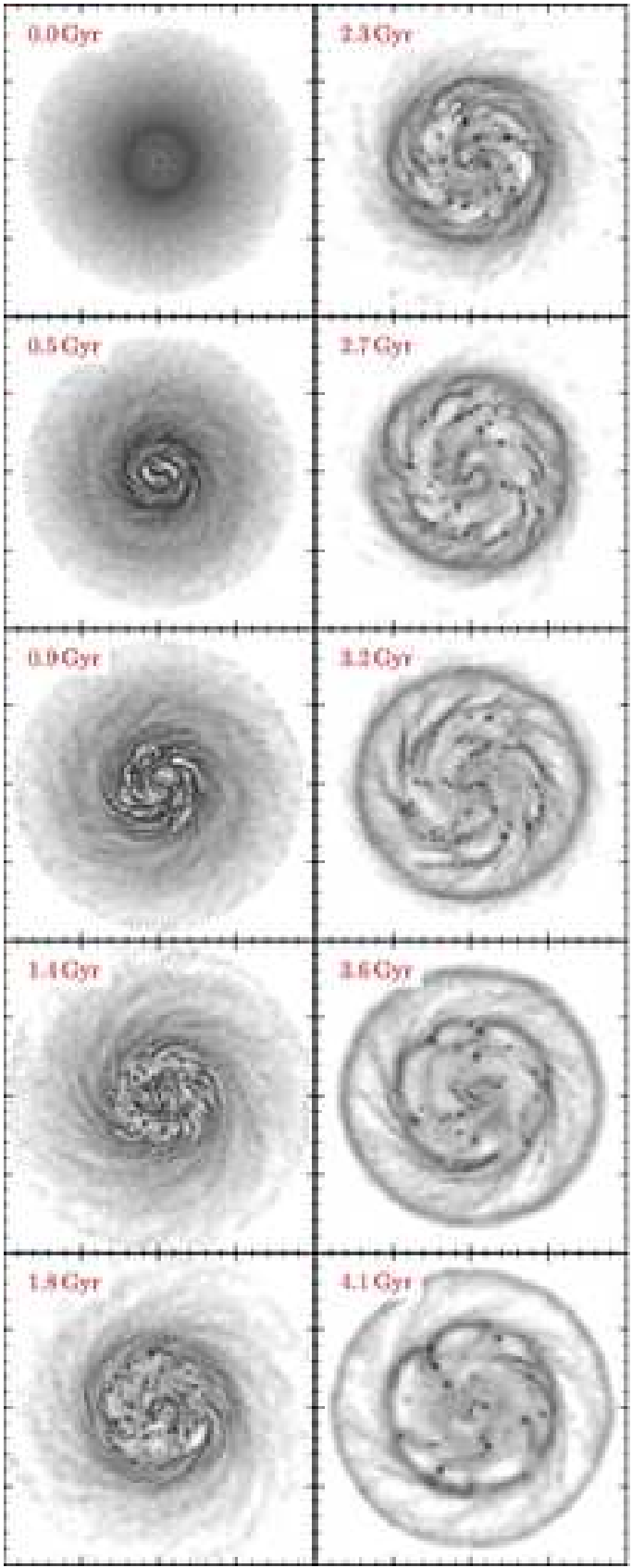}}
  \caption{Evolution of model $N=3'$ where the all the gas (visible and dark) is assumed
  to be dissipative. Each image represents
  the surface density of the visible gas. The box size is $200\times200\,\rm{kpc}$.}
  \label{evolution_000}
  \end{figure}
%

%

\end{appendix}

\bibliographystyle{aa}
\bibliography{bibliography}

\begin{thebibliography}{109}
\expandafter\ifx\csname natexlab\endcsname\relax\def\natexlab#1{#1}\fi

\bibitem[{{Appleton} {et~al.}(2006){Appleton}, {Xu}, {Reach}, {Dopita}, {Gao},
  {Lu}, {Popescu}, {Sulentic}, {Tuffs}, \& {Yun}}]{appleton06}
{Appleton}, P.~N., {Xu}, K.~C., {Reach}, W., {et~al.} 2006, \apjl, 639, L51

\bibitem[{{Athanassoula}(1984)}]{athanassoula84}
{Athanassoula}, E. 1984, \physrep, 114, 321

\bibitem[{{Begum} {et~al.}(2008){Begum}, {Chengalur}, {Karachentsev}, \&
  {Sharina}}]{begum08}
{Begum}, A., {Chengalur}, J.~N., {Karachentsev}, I.~D., \& {Sharina}, M.~E.
  2008, \mnras, 386, 138

\bibitem[{{Binney} \& {Tremaine}(1987)}]{binney87}
{Binney}, J. \& {Tremaine}, S. 1987, {Galactic dynamics} (Princeton, NJ,
  Princeton University Press, 1987)

\bibitem[{{Bissantz} \& {Gerhard}(2002)}]{bissantz02}
{Bissantz}, N. \& {Gerhard}, O. 2002, \mnras, 330, 591

\bibitem[{{Blais-Ouellette} {et~al.}(2001){Blais-Ouellette}, {Amram}, \&
  {Carignan}}]{blaisouellette01}
{Blais-Ouellette}, S., {Amram}, P., \& {Carignan}, C. 2001, \aj, 121, 1952

\bibitem[{{Boomsma} {et~al.}(2008){Boomsma}, {Oosterloo}, {Fraternali}, {van
  der Hulst}, \& {Sancisi}}]{boomsma08}
{Boomsma}, R., {Oosterloo}, T.~A., {Fraternali}, F., {van der Hulst}, J.~M., \&
  {Sancisi}, R. 2008, ArXiv e-prints, 807

\bibitem[{{Bosma}(1978)}]{bosma78}
{Bosma}, A. 1978, PhD thesis, PhD Thesis, Groningen Univ., (1978)

\bibitem[{{Bosma}(1981)}]{bosma81a}
{Bosma}, A. 1981, \aj, 86, 1825

\bibitem[{{Bottema}(2003)}]{bottema03}
{Bottema}, R. 2003, \mnras, 344, 358

\bibitem[{{Boulanger} {et~al.}(2008){Boulanger}, {Maillard}, {Appleton},
  {Falgarone}, {Lagache}, {Schulz}, {Wakker}, {Bressan}, {Cernicharo},
  {Charmandaris}, {Drissen}, {Helou}, {Henning}, {Lim}, {Valentjin}, {Abergel},
  {Bourlot}, {Bouzit}, {Cabrit}, {Combes}, {Deharveng}, {Desmet}, {Dole},
  {Dumesnil}, {Dutrey}, {Fourmond}, {Gavila}, {Grang{\'e}}, {Gry}, {Guillard},
  {Guilloteau}, {Habart}, {Huet}, {Joblin}, {Langer}, {Longval}, {Madden},
  {Martin}, {Miville-Desch{\^e}nes}, {Pineau Des For{\^e}ts}, {Pointecouteau},
  {Roussel}, {Tresse}, {Verstraete}, {Viallefond}, {Bertoldi}, {Jorgensen},
  {Bouwman}, {Carmona}, {Krause}, {Baruffolo}, {Bonoli}, {Bortoletto},
  {Danese}, {Granato}, {Pernechele}, {Rampazzo}, {Silva}, {Zotti}, {Pardo},
  {Spaans}, {van der Tak}, {Wild}, {Ferlet}, {Ramsay Howat}, {Smith},
  {Swinyard}, {Wright}, {Joncas}, {Martin}, {Davis}, {Draine}, {Goldsmith},
  {Mainzer}, {Ogle}, {Rinehart}, {Stacey}, \& {Tielens}}]{boulanger08}
{Boulanger}, F., {Maillard}, J.~P., {Appleton}, P., {et~al.} 2008, Experimental
  Astronomy, 20

\bibitem[{{Bournaud} \& {Combes}(2002)}]{bournaud02}
{Bournaud}, F. \& {Combes}, F. 2002, \aap, 392, 83

\bibitem[{{Bournaud} {et~al.}(2007){Bournaud}, {Duc}, {Brinks}, {Boquien},
  {Amram}, {Lisenfeld}, {Koribalski}, {Walter}, \& {Charmandaris}}]{bournaud07}
{Bournaud}, F., {Duc}, P.-A., {Brinks}, E., {et~al.} 2007, Science, 316, 1166

\bibitem[{{Brahic}(1977)}]{brahic77}
{Brahic}, A. 1977, \aap, 54, 895

\bibitem[{{Broeils}(1992)}]{broeils92}
{Broeils}, A.~H. 1992, PhD thesis, PhD thesis, Univ.~Groningen, (1992)

\bibitem[{{Bureau} {et~al.}(1999){Bureau}, {Freeman}, {Pfitzner}, \&
  {Meurer}}]{bureau99}
{Bureau}, M., {Freeman}, K.~C., {Pfitzner}, D.~W., \& {Meurer}, G.~R. 1999,
  \aj, 118, 2158

\bibitem[{{Carignan} {et~al.}(1990){Carignan}, {Charbonneau}, {Boulanger}, \&
  {Viallefond}}]{carignan90}
{Carignan}, C., {Charbonneau}, P., {Boulanger}, F., \& {Viallefond}, F. 1990,
  \aap, 234, 43

\bibitem[{{Cen} \& {Ostriker}(1999)}]{cen99}
{Cen}, R. \& {Ostriker}, J.~P. 1999, \apj, 514, 1

\bibitem[{{Cen} \& {Ostriker}(2006)}]{cen06}
{Cen}, R. \& {Ostriker}, J.~P. 2006, \apj, 650, 560

\bibitem[{{Combes} \& {Gerin}(1985)}]{combes84}
{Combes}, F. \& {Gerin}, M. 1985, \aap, 150, 327

\bibitem[{{Combes} \& {Pfenniger}(1997)}]{combes97}
{Combes}, F. \& {Pfenniger}, D. 1997, \aap, 327, 453

\bibitem[{{Crosthwaite} {et~al.}(2002){Crosthwaite}, {Turner}, {Buchholz},
  {Ho}, \& {Martin}}]{crosthwaite02}
{Crosthwaite}, L.~P., {Turner}, J.~L., {Buchholz}, L., {Ho}, P.~T.~P., \&
  {Martin}, R.~N. 2002, \aj, 123, 1892

\bibitem[{{Cuillandre} {et~al.}(2001){Cuillandre}, {Lequeux}, {Allen},
  {Mellier}, \& {Bertin}}]{cuillandre01}
{Cuillandre}, J.-C., {Lequeux}, J., {Allen}, R.~J., {Mellier}, Y., \& {Bertin},
  E. 2001, \apj, 554, 190

\bibitem[{{de Blok}(2005)}]{deblock05}
{de Blok}, W.~J.~G. 2005, \apj, 634, 227

\bibitem[{{de Blok} \& {Bosma}(2002)}]{deblok02}
{de Blok}, W.~J.~G. \& {Bosma}, A. 2002, \aap, 385, 816

\bibitem[{{de Blok} \& {Walter}(2003)}]{deblok03}
{de Blok}, W.~J.~G. \& {Walter}, F. 2003, \mnras, 341, L39

\bibitem[{{de Blok} {et~al.}(2008){de Blok}, {Walter}, {Brinks},
  {Trachternach}, {Oh}, \& {Kennicutt}}]{deblok08}
{de Blok}, W.~J.~G., {Walter}, F., {Brinks}, E., {et~al.} 2008, \aj, 136, 2648

\bibitem[{{Draine}(1978)}]{draine78}
{Draine}, B.~T. 1978, \apjs, 36, 595

\bibitem[{{Draine}(1998)}]{draine98}
{Draine}, B.~T. 1998, \apjl, 509, L41

\bibitem[{{Dubinski} {et~al.}(2008){Dubinski}, {Gauthier}, {Widrow}, \&
  {Nickerson}}]{dubinski08}
{Dubinski}, J., {Gauthier}, J.-R., {Widrow}, L., \& {Nickerson}, S. 2008, in
  Astronomical Society of the Pacific Conference Series, Vol. 396, Astronomical
  Society of the Pacific Conference Series, ed. J.~G. {Funes} \& E.~M.
  {Corsini}, 321

\bibitem[{{Egami} {et~al.}(2006){Egami}, {Rieke}, {Fadda}, \&
  {Hines}}]{egami06}
{Egami}, E., {Rieke}, G.~H., {Fadda}, D., \& {Hines}, D.~C. 2006, \apjl, 652,
  L21

\bibitem[{{Elmegreen}(1995)}]{elmegreen95}
{Elmegreen}, B.~G. 1995, \mnras, 275, 944

\bibitem[{{Elmegreen}(1997)}]{elmegreen97}
{Elmegreen}, B.~G. 1997, in New extragalactic perspectives in the new South
  Africa / Kluwer, 1996, 117, 467

\bibitem[{{Ferguson} {et~al.}(1998){Ferguson}, {Wyse}, {Gallagher}, \&
  {Hunter}}]{ferguson98}
{Ferguson}, A.~M.~N., {Wyse}, R.~F.~G., {Gallagher}, J.~S., \& {Hunter}, D.~A.
  1998, \apjl, 506, L19

\bibitem[{{Font} {et~al.}(2001){Font}, {Navarro}, {Stadel}, \&
  {Quinn}}]{font01}
{Font}, A.~S., {Navarro}, J.~F., {Stadel}, J., \& {Quinn}, T. 2001, \apjl, 563,
  L1

\bibitem[{{Fukugita} \& {Peebles}(2004)}]{fukugita04}
{Fukugita}, M. \& {Peebles}, P.~J.~E. 2004, \apj, 616, 643

\bibitem[{{Fux}(2005)}]{fux05}
{Fux}, R. 2005, \aap, 430, 853

\bibitem[{{Gao} {et~al.}(2004){Gao}, {White}, {Jenkins}, {Stoehr}, \&
  {Springel}}]{gao04}
{Gao}, L., {White}, S.~D.~M., {Jenkins}, A., {Stoehr}, F., \& {Springel}, V.
  2004, \mnras, 355, 819

\bibitem[{{Gauthier} {et~al.}(2006){Gauthier}, {Dubinski}, \&
  {Widrow}}]{gauthier06}
{Gauthier}, J.-R., {Dubinski}, J., \& {Widrow}, L.~M. 2006, \apj, 653, 1180

\bibitem[{{Gentile} {et~al.}(2005){Gentile}, {Burkert}, {Salucci}, {Klein}, \&
  {Walter}}]{gentile05}
{Gentile}, G., {Burkert}, A., {Salucci}, P., {Klein}, U., \& {Walter}, F. 2005,
  \apjl, 634, L145

\bibitem[{{Gentile} {et~al.}(2004){Gentile}, {Salucci}, {Klein}, {Vergani}, \&
  {Kalberla}}]{gentile04}
{Gentile}, G., {Salucci}, P., {Klein}, U., {Vergani}, D., \& {Kalberla}, P.
  2004, \mnras, 351, 903

\bibitem[{{Gerhard}(2006)}]{gerhard06}
{Gerhard}, O. 2006, in EAS Publications Series, Vol.~20, EAS Publications
  Series, ed. G.~A. {Mamon}, F.~{Combes}, C.~{Deffayet}, \& B.~{Fort}, 89--96

\bibitem[{{Gerhard} \& {Silk}(1996)}]{gerhard96}
{Gerhard}, O. \& {Silk}, J. 1996, \apj, 472, 34

\bibitem[{{Goldreich} \& {Lynden-Bell}(1965)}]{goldreich65}
{Goldreich}, P. \& {Lynden-Bell}, D. 1965, \mnras, 130, 125

\bibitem[{{Goldsmith} {et~al.}(1969){Goldsmith}, {Habing}, \&
  {Field}}]{goldsmith69}
{Goldsmith}, D.~W., {Habing}, H.~J., \& {Field}, G.~B. 1969, \apj, 158, 173

\bibitem[{{Grenier} {et~al.}(2005){Grenier}, {Casandjian}, \&
  {Terrier}}]{grenier05}
{Grenier}, I.~A., {Casandjian}, J.-M., \& {Terrier}, R. 2005, Science, 307,
  1292

\bibitem[{{Habing}(1968)}]{habing68}
{Habing}, H.~J. 1968, \bain, 19, 421

\bibitem[{{Harfst} {et~al.}(2006){Harfst}, {Theis}, \& {Hensler}}]{harfst06}
{Harfst}, S., {Theis}, C., \& {Hensler}, G. 2006, \aap, 449, 509

\bibitem[{{Heithausen}(2002)}]{heithausen02}
{Heithausen}, A. 2002, \aap, 393, L41

\bibitem[{{Heithausen}(2004)}]{heithausen04}
{Heithausen}, A. 2004, \apjl, 606, L13

\bibitem[{{Henriksen} \& {Widrow}(1995)}]{henriksen95}
{Henriksen}, R.~N. \& {Widrow}, L.~M. 1995, \apj, 441, 70

\bibitem[{{Hernquist}(1993)}]{hernquist93}
{Hernquist}, L. 1993, \apjs, 86, 389

\bibitem[{{Hoekstra} {et~al.}(2001){Hoekstra}, {van Albada}, \&
  {Sancisi}}]{hoekstra01}
{Hoekstra}, H., {van Albada}, T.~S., \& {Sancisi}, R. 2001, \mnras, 323, 453

\bibitem[{{Jiang} \& {Binney}(1999)}]{jiang99}
{Jiang}, I.-G. \& {Binney}, J. 1999, \mnras, 303, L7

\bibitem[{{Jog}(1996)}]{jog96}
{Jog}, C.~J. 1996, \mnras, 278, 209

\bibitem[{{Jog} \& {Solomon}(1984{\natexlab{a}})}]{jog84b}
{Jog}, C.~J. \& {Solomon}, P.~M. 1984{\natexlab{a}}, \apj, 276, 127

\bibitem[{{Jog} \& {Solomon}(1984{\natexlab{b}})}]{jog84a}
{Jog}, C.~J. \& {Solomon}, P.~M. 1984{\natexlab{b}}, \apj, 276, 114

\bibitem[{{Johnston} {et~al.}(2005){Johnston}, {Law}, \&
  {Majewski}}]{johnston05}
{Johnston}, K.~V., {Law}, D.~R., \& {Majewski}, S.~R. 2005, \apj, 619, 800

\bibitem[{{Julian} \& {Toomre}(1966)}]{julian66}
{Julian}, W.~H. \& {Toomre}, A. 1966, \apj, 146, 810

\bibitem[{{Kalberla}(2003)}]{kalberla03}
{Kalberla}, P.~M.~W. 2003, \apj, 588, 805

\bibitem[{{Kalberla}(2004)}]{kalberla04}
{Kalberla}, P.~M.~W. 2004, \apss, 289, 239

\bibitem[{{Kalberla} {et~al.}(2007){Kalberla}, {Dedes}, {Kerp}, \&
  {Haud}}]{kalberla07}
{Kalberla}, P.~M.~W., {Dedes}, L., {Kerp}, J., \& {Haud}, U. 2007, \aap, 469,
  511

\bibitem[{{Kalberla} \& {Kerp}(1998)}]{kalberla98}
{Kalberla}, P.~M.~W. \& {Kerp}, J. 1998, \aap, 339, 745

\bibitem[{{Katz} {et~al.}(1996){Katz}, {Weinberg}, \& {Hernquist}}]{katz96}
{Katz}, N., {Weinberg}, D.~H., \& {Hernquist}, L. 1996, \apjs, 105, 19

\bibitem[{{Kaufmann} {et~al.}(2007){Kaufmann}, {Mayer}, {Wadsley}, {Stadel}, \&
  {Moore}}]{kaufmann07}
{Kaufmann}, T., {Mayer}, L., {Wadsley}, J., {Stadel}, J., \& {Moore}, B. 2007,
  \mnras, 375, 53

\bibitem[{{Kere{\v s}} {et~al.}(2005){Kere{\v s}}, {Katz}, {Weinberg}, \&
  {Dav{\'e}}}]{keres05}
{Kere{\v s}}, D., {Katz}, N., {Weinberg}, D.~H., \& {Dav{\'e}}, R. 2005,
  \mnras, 363, 2

\bibitem[{{Kerins} {et~al.}(2002){Kerins}, {Binney}, \& {Silk}}]{kerins02}
{Kerins}, E., {Binney}, J., \& {Silk}, J. 2002, \mnras, 332, L29

\bibitem[{{Klypin} {et~al.}(1999){Klypin}, {Kravtsov}, {Valenzuela}, \&
  {Prada}}]{klypin99}
{Klypin}, A., {Kravtsov}, A.~V., {Valenzuela}, O., \& {Prada}, F. 1999, \apj,
  522, 82

\bibitem[{{Leli{\`e}vre} \& {Roy}(2000)}]{lelievre00}
{Leli{\`e}vre}, M. \& {Roy}, J.-R. 2000, \aj, 120, 1306

\bibitem[{{Maio} {et~al.}(2007){Maio}, {Dolag}, {Ciardi}, \&
  {Tornatore}}]{maio07}
{Maio}, U., {Dolag}, K., {Ciardi}, B., \& {Tornatore}, L. 2007, \mnras, 379,
  963

\bibitem[{{Masset} \& {Bureau}(2003)}]{masset03}
{Masset}, F.~S. \& {Bureau}, M. 2003, \apj, 586, 152

\bibitem[{{Mayer} {et~al.}(2008){Mayer}, {Governato}, \& {Kaufmann}}]{mayer08}
{Mayer}, L., {Governato}, F., \& {Kaufmann}, T. 2008, Advanced Science Letters,
  1, 7

\bibitem[{{Mayer} \& {Moore}(2004)}]{mayer04}
{Mayer}, L. \& {Moore}, B. 2004, \mnras, 354, 477

\bibitem[{{Merlin} \& {Chiosi}(2007)}]{merlin07}
{Merlin}, E. \& {Chiosi}, C. 2007, \aap, 473, 733

\bibitem[{{Miyamoto} \& {Nagai}(1975)}]{miyamoto75}
{Miyamoto}, M. \& {Nagai}, R. 1975, \pasj, 27, 533

\bibitem[{{Moore} {et~al.}(1999){Moore}, {Ghigna}, {Governato}, {Lake},
  {Quinn}, {Stadel}, \& {Tozzi}}]{moore99}
{Moore}, B., {Ghigna}, S., {Governato}, F., {et~al.} 1999, \apjl, 524, L19

\bibitem[{{Navarro} \& {Benz}(1991)}]{navarro91}
{Navarro}, J.~F. \& {Benz}, W. 1991, \apj, 380, 320

\bibitem[{{Navarro} \& {Steinmetz}(1997)}]{navarro97}
{Navarro}, J.~F. \& {Steinmetz}, M. 1997, \apj, 478, 13

\bibitem[{{Ogle} {et~al.}(2007){Ogle}, {Antonucci}, {Appleton}, \&
  {Whysong}}]{ogle07}
{Ogle}, P., {Antonucci}, R., {Appleton}, P.~N., \& {Whysong}, D. 2007, \apj,
  668, 699

\bibitem[{{Papadopoulos} {et~al.}(2002){Papadopoulos}, {Thi}, \&
  {Viti}}]{papadopoulos02}
{Papadopoulos}, P.~P., {Thi}, W.-F., \& {Viti}, S. 2002, \apj, 579, 270

\bibitem[{{Pfenniger}(2004)}]{pfenniger04}
{Pfenniger}, D. 2004, in Baryons in Dark Matter Halos, ed. R.~{Dettmar},
  U.~{Klein}, \& P.~{Salucci}

\bibitem[{{Pfenniger} \& {Combes}(1994)}]{pfenniger94b}
{Pfenniger}, D. \& {Combes}, F. 1994, \aap, 285, 94

\bibitem[{{Pfenniger} {et~al.}(1994){Pfenniger}, {Combes}, \&
  {Martinet}}]{pfenniger94a}
{Pfenniger}, D., {Combes}, F., \& {Martinet}, L. 1994, \aap, 285, 79

\bibitem[{{Pfenniger} \& {Revaz}(2005)}]{pfenniger05}
{Pfenniger}, D. \& {Revaz}, Y. 2005, \aap, 431, 511

\bibitem[{{Reach} {et~al.}(1995){Reach}, {Franz}, {Weiland}, {Hauser},
  {Kelsall}, {Wright}, {Rawley}, {Stemwedel}, \& {Spiesman}}]{reach95}
{Reach}, W.~T., {Franz}, B.~A., {Weiland}, J.~L., {et~al.} 1995, \nat, 374, 521

\bibitem[{{Reshetnikov} \& {Combes}(1999)}]{reshetnikov99}
{Reshetnikov}, V. \& {Combes}, F. 1999, \aaps, 138, 101

\bibitem[{{Revaz} \& {Pfenniger}(2004)}]{revaz04}
{Revaz}, Y. \& {Pfenniger}, D. 2004, \aap, 425, 67

\bibitem[{{Revaz} \& {Pfenniger}(2007)}]{revaz07}
{Revaz}, Y. \& {Pfenniger}, D. 2007, {a New Scenario for the Origin of
  Galacticwarps} (Island Universes - Structure and Evolution of Disk Galaxies),
  149

\bibitem[{{Safa} \& {Pfenniger}(2008)}]{safa08}
{Safa}, Y. \& {Pfenniger}, D. 2008, European Physical Journal B, 66, 337

\bibitem[{{S{\'a}nchez-Saavedra} {et~al.}(2003){S{\'a}nchez-Saavedra},
  {Battaner}, {Guijarro}, {L{\'o}pez-Corredoira}, \&
  {Castro-Rodr{\'{\i}}guez}}]{sanchez-saavedra03}
{S{\'a}nchez-Saavedra}, M.~L., {Battaner}, E., {Guijarro}, A.,
  {L{\'o}pez-Corredoira}, M., \& {Castro-Rodr{\'{\i}}guez}, N. 2003, \aap, 399,
  457

\bibitem[{{Schwarz}(1981)}]{schwarz81}
{Schwarz}, M.~P. 1981, \apj, 247, 77

\bibitem[{{Seiden} {et~al.}(1984){Seiden}, {Schulman}, \&
  {Elmegreen}}]{seiden84}
{Seiden}, P.~E., {Schulman}, L.~S., \& {Elmegreen}, B.~G. 1984, \apj, 282, 95

\bibitem[{{Sellwood} \& {Binney}(2002)}]{sellwood02}
{Sellwood}, J.~A. \& {Binney}, J.~J. 2002, \mnras, 336, 785

\bibitem[{{Semelin} \& {Combes}(2002)}]{semelin02}
{Semelin}, B. \& {Combes}, F. 2002, \aap, 388, 826

\bibitem[{{Shen} \& {Sellwood}(2006)}]{shen06}
{Shen}, J. \& {Sellwood}, J.~A. 2006, \mnras, 370, 2

\bibitem[{{Smith} {et~al.}(2000){Smith}, {Allen}, {Bohlin}, {Nicholson}, \&
  {Stecher}}]{smith00}
{Smith}, D.~A., {Allen}, R.~J., {Bohlin}, R.~C., {Nicholson}, N., \& {Stecher},
  T.~P. 2000, \apj, 538, 608

\bibitem[{{Spano} {et~al.}(2008){Spano}, {Marcelin}, {Amram}, {Carignan},
  {Epinat}, \& {Hernandez}}]{spano08}
{Spano}, M., {Marcelin}, M., {Amram}, P., {et~al.} 2008, \mnras, 383, 297

\bibitem[{{Spekkens} {et~al.}(2005){Spekkens}, {Giovanelli}, \&
  {Haynes}}]{spekkens05}
{Spekkens}, K., {Giovanelli}, R., \& {Haynes}, M.~P. 2005, \aj, 129, 2119

\bibitem[{{Springel}(2005)}]{springel05}
{Springel}, V. 2005, \mnras, 364, 1105

\bibitem[{{Springel} {et~al.}(2006){Springel}, {Frenk}, \&
  {White}}]{springel06}
{Springel}, V., {Frenk}, C.~S., \& {White}, S.~D.~M. 2006, \nat, 440, 1137

\bibitem[{{Strigari} {et~al.}(2007){Strigari}, {Bullock}, {Kaplinghat},
  {Diemand}, {Kuhlen}, \& {Madau}}]{stringari07}
{Strigari}, L.~E., {Bullock}, J.~S., {Kaplinghat}, M., {et~al.} 2007, \apj,
  669, 676

\bibitem[{{Swaters} {et~al.}(2003){Swaters}, {Madore}, {van den Bosch}, \&
  {Balcells}}]{swaters03}
{Swaters}, R.~A., {Madore}, B.~F., {van den Bosch}, F.~C., \& {Balcells}, M.
  2003, \apj, 583, 732

\bibitem[{{Thilker} {et~al.}(2005){Thilker}, {Bianchi}, {Boissier}, {Gil de
  Paz}, {Madore}, {Martin}, {Meurer}, {Neff}, {Rich}, {Schiminovich},
  {Seibert}, {Wyder}, {Barlow}, {Byun}, {Donas}, {Forster}, {Friedman},
  {Heckman}, {Jelinsky}, {Lee}, {Malina}, {Milliard}, {Morrissey}, {Siegmund},
  {Small}, {Szalay}, \& {Welsh}}]{thilker05}
{Thilker}, D.~A., {Bianchi}, L., {Boissier}, S., {et~al.} 2005, \apjl, 619, L79

\bibitem[{{Toomre}(1981)}]{toomre81}
{Toomre}, A. 1981, in Structure and Evolution of Normal Galaxies, ed. S.~M.
  {Fall} \& D.~{Lynden-Bell}, 111--136

\bibitem[{{van den Bosch}(2001)}]{van_den_bosch01}
{van den Bosch}, F.~C. 2001, \mnras, 327, 1334

\bibitem[{{Walker} \& {Wardle}(1998)}]{walker98}
{Walker}, M. \& {Wardle}, M. 1998, \apjl, 498, L125+

\bibitem[{{Wang} \& {Silk}(1994)}]{wang94}
{Wang}, B. \& {Silk}, J. 1994, \apj, 427, 759

\bibitem[{{Watson}(1972)}]{watson72}
{Watson}, W.~D. 1972, \apj, 176, 103

\bibitem[{{Weinberg} \& {Blitz}(2006)}]{weinberg06}
{Weinberg}, M.~D. \& {Blitz}, L. 2006, \apjl, 641, L33

\end{thebibliography}

\end{document}